\documentclass[10pt,conference]{IEEEtran}
\usepackage{cite}
\usepackage{amsmath,amssymb,amsfonts}
\usepackage{algorithmic}
\usepackage{graphicx}
\usepackage{textcomp}
\usepackage{xcolor}
\def\BibTeX{{\rm B\kern-.05em{\sc i\kern-.025em b}\kern-.08em
    T\kern-.1667em\lower.7ex\hbox{E}\kern-.125emX}}

\usepackage{braket}
\usepackage{amsmath}
\usepackage{mathtools}
\usepackage{courier}
\usepackage{tikz}
\usetikzlibrary{quantikz}

\usepackage{lipsum}
\usepackage{afterpage}
\usepackage{stfloats}
\usepackage{caption}
\usepackage{subcaption}

\DeclareMathOperator{\diag}{diag}
\DeclareMathOperator{\sign}{sign}
\DeclareMathOperator{\qubits}{qubits}

\DeclareMathOperator{\Tr}{Tr}
\DeclarePairedDelimiter{\abs}{\lvert}{\rvert}

\newcommand{\passed}{\mathrm{passed}}
\newcommand{\caught}{\mathrm{caught}}
\newcommand{\remaining}{\mathrm{remaining}}

\newcommand{\tmax}{\mathrm{max}}

\usepackage{float}  % allow [H] figure placement
\usepackage{graphicx}
\graphicspath{{./figures/}} % set path for all figures

\usepackage{enumitem}

\usepackage[linesnumbered,ruled,vlined]{algorithm2e}
\SetAlgoCaptionSeparator{:}
\SetAlCapFnt{\normalsize}
\SetAlCapNameFnt{\normalsize}

\begin{document}

\title{Circuit decompositions and scheduling for neutral atom devices with limited local addressability}

\author{
\IEEEauthorblockN{Natalia Nottingham}
\IEEEauthorblockA{\textit{Dept.~of Computer Science} \\
\textit{University of Chicago}\\
Chicago, IL, USA \\
nottingham@uchicago.edu}
\and
\IEEEauthorblockN{Michael A.~Perlin}
\IEEEauthorblockA{\textit{Global Technology Applied Research} \\
\textit{JPMorganChase} \\
New York, NY, USA \\
michael.perlin@jpmchase.com}
\and
\IEEEauthorblockN{Dhirpal Shah}
\IEEEauthorblockA{\textit{Dept.~of Computer Science} \\
\textit{University of Chicago}\\
Chicago, IL, USA \\
dhirpalshah@uchicago.edu}
\and
\IEEEauthorblockN{Ryan White}
\IEEEauthorblockA{\textit{Dept.~of Physics} \\
\textit{University of Chicago}\\
Chicago, IL, USA \\
rpwhite@uchicago.edu}
\and
\IEEEauthorblockN{Hannes Bernien}
\IEEEauthorblockA{\textit{Pritzker School of Molecular Engineering} \\
\textit{University of Chicago}\\
Chicago, IL, USA \\
bernien@uchicago.edu}
\and
\IEEEauthorblockN{Frederic T.~Chong}
\IEEEauthorblockA{\textit{Dept.~of Computer Science} \\
\textit{University of Chicago}\\
Chicago, IL, USA \\
chong@cs.uchicago.edu}
\and
\IEEEauthorblockN{Jonathan M.~Baker}
\IEEEauthorblockA{\textit{Dept.~Electrical \& Computer Engineering} \\
\textit{University of Texas at Austin}\\
Austin, TX, USA \\
jonathan.baker@austin.utexas.edu}
}

\maketitle

\begin{abstract}

Despite major ongoing advancements in neutral atom hardware technology, there remains limited work in systems-level software tailored to overcoming the challenges of neutral atom quantum computers. In particular, most current neutral atom architectures do not natively support local addressing of single-qubit rotations about an axis in the xy-plane of the Bloch sphere. Instead, these are executed via global beams applied simultaneously to all qubits. While previous neutral atom experimental work has used straightforward synthesis methods to convert short sequences of operations into this native gate set, these methods cannot be incorporated into a systems-level framework nor applied to entire circuits without imposing impractical amounts of serialization. Without sufficient compiler optimizations, decompositions involving global gates will significantly increase circuit depth, gate count, and accumulation of errors. No prior compiler work has addressed this, and adapting existing compilers to solve this problem is nontrivial.

In this paper, we present an optimized compiler pipeline that translates an input circuit from an arbitrary gate set into a realistic neutral atom native gate set containing global gates. We focus on decomposition and scheduling passes that minimize the final circuit's global gate count and total global rotation amount. As we show, these costs contribute the most to the circuit's duration and overall error, relative to costs incurred by other gate types. Compared to the unoptimized version of our compiler pipeline, minimizing global gate costs gives up to 4.77x speedup in circuit duration. Compared to the closest prior existing work, we achieve up to 53.8x speedup. For large circuits, we observe a few orders of magnitude improvement in circuit fidelities.

\end{abstract}


\section{Introduction}

Recent years have shown drastic improvements in the development of quantum computing hardware technologies, including superconducting qubits, neutral atoms, and trapped ions. Neutral atom quantum computers have demonstrated immense promise due to their exceptionally long coherence times, scalability, native multi-qubit gates, higher connectivity resulting from longer-range interactions, and the ability to produce identical and well-characterized qubits \cite{jaksch_fast_2000,morgado_quantum_2021}.

Despite these advancements, quantum computing remains in an era defined by large amounts of noise, high gate error rates, and the possibility of qubit states decohering prior to completion of the circuit. Given these constraints, hardware-aware optimizations at the compiler level---that successfully exploit the hardware platform's advantages while employing techniques to overcome its limitations---are essential. 

While considerable prior work exists on quantum compilation \cite{anis_qiskit_2021,cirq_developers_2022_7465577,sivarajah_tket_2020,cowtan_qubit_2019,murali_noise-adaptive_2019,li_tackling_2019}, the majority of such work has either been hardware-agnostic or tailored to the constraints of superconducting qubits, with limited work focusing on compiler approaches for neutral atoms. Baker et al.~\cite{baker_exploiting_2021}, Brandhofer et al.~\cite{brandhofer_optimal_2021}, Patel et al.~\cite{patel_geyser_2022}, Tan et al.~\cite{tan_qubit_2022}, and Li et al.~\cite{li_timing-aware_2023} took the first steps towards developing compilers specific to neutral atoms, accounting for atom loss, long-distance interactions, atom movement (shuttling), and native multi-qubit gates. 

However, these prior compiler works assume a native gate set in which all single-qubit gates can be executed via locally-addressing beams. This assumption is inconsistent with most current neutral atom architectures, in which the execution of some single-qubit gates is only supported natively via globally-addressing beams that rotate all qubits homogeneously \cite{graham_multi-qubit_2022}. Though global addressing presents fewer engineering challenges and lower costs on a hardware level, it creates a more difficult compiler problem: if a global gate is used to execute an operation that acts only on a small subset of qubits in the original circuit, the compiler must ensure that any operation on off-target qubits is ``undone''. This leads to increased circuit depths and gate counts which, without effective compiler optimizations, can significantly decrease fidelities and cause circuit runtimes to potentially exceed device coherence times.

In this paper, we present the first systems-level work to address this problem. In addition to outputting a circuit that is directly executable on neutral atom architectures with limited local addressability, our compiler minimizes gate counts and rotation angles of global gates; as we show, these dominate time and fidelity costs relative to other gate types. Achieving this objective relies on two key insights. First, by carefully choosing the rotation \textit{axis} of each global gate, we can reduce the rotation \textit{angle} required to decompose sets of parallel single-qubit gates. Furthermore, by choosing the same axis and angle to decompose single-qubit gates within the same moment, we can avoid sacrificing parallelism. Second, how we group single-qubit gates into moments impacts the range of global gate angles that will provide a valid decomposition---i.e., our \textit{scheduling} approach determines how much benefit is possible from our \textit{decomposition} methods. We apply these observations in the following contributions: 

\begin{figure}
    \centering
    \includegraphics[scale=0.49]{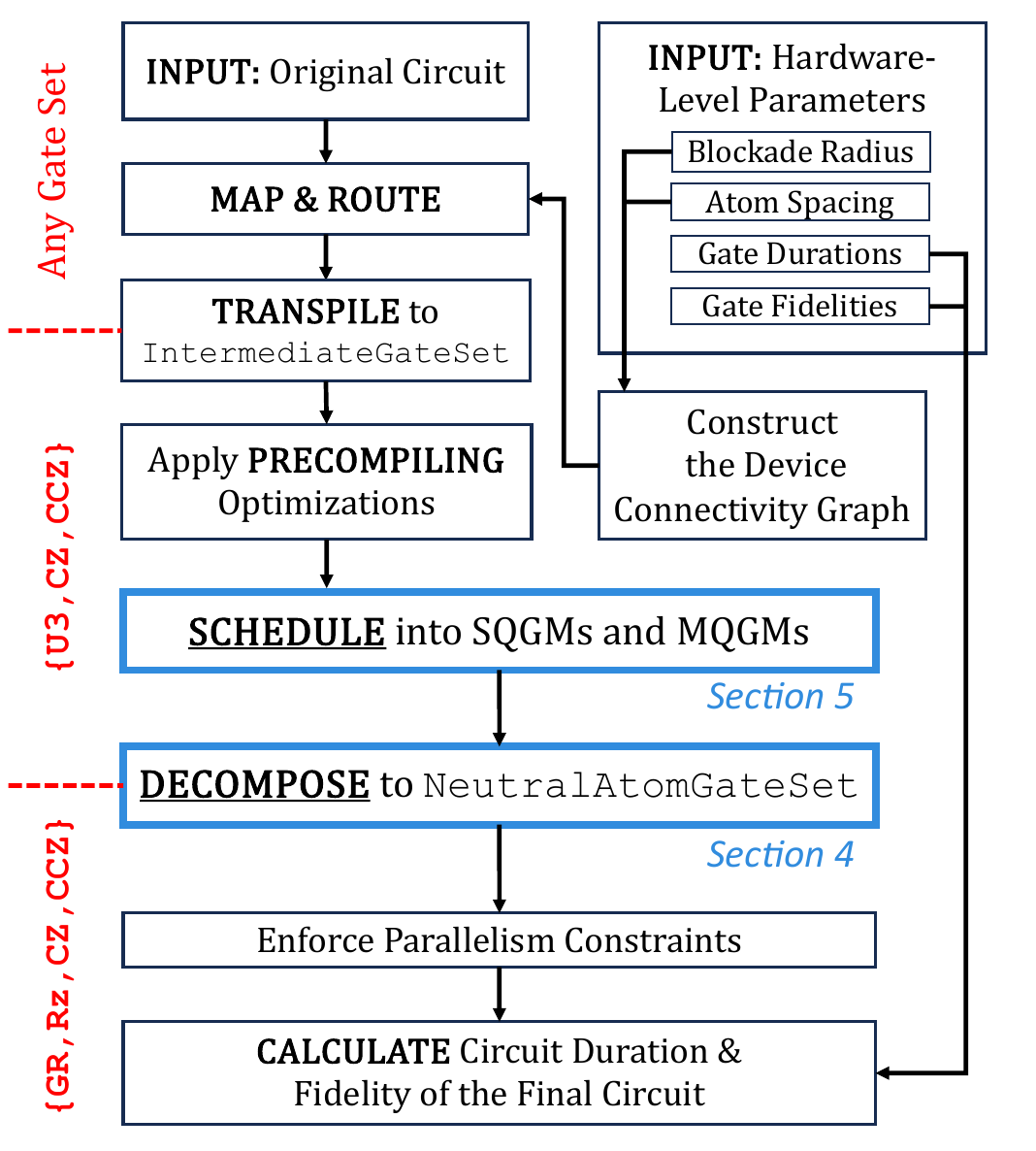}
    \caption{Overview of our compiler pipeline. In this work, we focus mainly on optimizing the steps to schedule the circuit (Sec.~\ref{section: scheduling}) and decompose into the \texttt{NeutralAtomGateSet} containing gates $\set{\texttt{GR},\texttt{Rz},\texttt{CZ},\texttt{CCZ}}$ (Sec.~\ref{section: decomposition}).}
    \label{fig: compiler}
\end{figure}

\begin{itemize}
    \item \textbf{Decomposition strategies} that translate an arbitrary input circuit into an equivalent circuit expressed in terms of a realistic neutral atom native gate set, in which \texttt{Rx} and \texttt{Ry} gates are implemented globally. For a given schedule, our compiler finds the optimal decomposition with respect to global rotation angle. 
    \item A \textbf{scheduling algorithm} that specifically accounts for our decomposition equations, arranging gates into moments such that the total global rotation amount and gate count in the final circuit (once decomposed) is minimized. 
    \item A \textbf{holistic compiler pipeline} (Fig.~\ref{fig: compiler}), which incorporates our support for global rotations with other features specific to neutral atom hardware, such as long-range connectivity and native multi-qubit gates.
    \item Our code is \textbf{open source} and can be found at the link https://github.com/natalianottingham/globalgatecompiler.
\end{itemize}

\section{Background}

\subsection{Native Gates \& Global Addressing}

High-level quantum programs typically contain gates that are not a part of the hardware's \textit{native gate set}, i.e., the set of gates that can be directly executed on the hardware. A crucial role of the compiler is to translate the input quantum circuit into an equivalent circuit using only gates in the native set.

With many quantum hardware platforms, the compiler only needs to check if a particular \textit{operation} is executable on the hardware. With neutral atoms, however, the compiler must also consider whether the \textit{addressability} aligns with the architecture's capabilities. Specifically, most current neutral atom technology natively supports globally-addressing single-qubit gates about any axis in the Bloch sphere's xy-plane, but not locally-addressing gates of the same operation. Here, \textit{locally-addressing} or \textit{individually-addressing} gates refer to those that can be applied to a chosen qubit without affecting any other qubit states. In contrast, \textit{globally-addressing} gates simultaneously apply the same operation (in terms of both rotation \textit{axis} and \textit{angle}) to all qubits in the circuit. 

The neutral atom native gate set we consider, referred to as the \texttt{NeutralAtomGateSet},
consists of local \texttt{CZ} gates, local \texttt{CCZ} gates, local \texttt{Rz}$(\lambda)$ gates, and global \texttt{GR}$(\theta,\phi)$ gates:
\begin{gather}
    \texttt{CZ} = \diag(1, 1, 1, -1),
    \label{eqn:CZ_gate_matrix} \\
    \texttt{CCZ} = \diag(1, 1, 1, 1, 1, 1, 1, -1),\label{eqn:CCZ_gate_matrix} \\
    \texttt{Rz}(\lambda) = \exp\left(-i\lambda \hat{Z}/2\right)
    = \diag(e^{-i\lambda/2}, e^{i\lambda/2}),
    \label{eqn:Rz_gate_matrix} \\
    \texttt{GR}(\theta,\phi)
    = \exp\left(-i\frac{\theta}{2}\sum_{j=1}^{n} (\cos(\phi)\hat{X_j} + \sin(\phi)\hat{Y_j})\right).
    \label{eqn:GR_gate_matrix}
\end{gather}

Here, $\hat{Z}$ is a single-qubit Pauli-$Z$ matrix; $\hat{X}_j$ and $\hat{Y}_j$ are Pauli-$X$ and Pauli-$Y$ matrices acting on qubit $j$; $\lambda$ and $\theta$ are rotation angles for \texttt{Rz} and \texttt{GR} gates, respectively; and the angle $\phi$ parameterizes the \texttt{GR} gate's axis of rotation. The globally-addressing \texttt{GR}$(\theta,\phi)$ gate implements the same operation as $n$ locally-addressing \texttt{R}$(\theta,\phi)=\exp(-i\frac{\theta}{2}(\cos(\phi)\hat X+\sin(\phi)\hat Y))$ gates executed separately on every qubit in the circuit.

Though not included in the final circuit's gate set, intermediate steps in our compiler also use the \texttt{U3} gate, which defines any arbitrary single-qubit rotation in terms of the Euler-angle parameters $\theta$, $\phi$, and $\lambda$:
\begin{equation}
    \texttt{U3}(\theta,\phi,\lambda) = \begin{pmatrix}
        \mathrm{cos}(\theta/2) & -e^{i\lambda}\mathrm{sin}(\theta/2) \\
        e^{i\phi}\mathrm{sin}(\theta/2) & e^{i(\phi+\lambda)}\mathrm{cos}(\theta/2)
    \end{pmatrix}.
\end{equation}
%\begin{equation}
%    \texttt{U3}(\theta,\phi,\lambda) = e^{i(\phi+\lambda)/2} \texttt{Rz}(\phi) \texttt{Ry}(\theta) \texttt{Rz}(\lambda),
%\end{equation}
%where $\texttt{Ry}(\theta) = \texttt{R}(\theta,\frac{\pi}{2})$.

\subsection{Neutral Atom Hardware}\label{section: hardware background}

With neutral atom quantum computing, the computational states $\ket{0}$ and $\ket{1}$ are typically encoded in hyperfine ground states of an alkali atom (such as Cesium or Rubidium), or alkaline-earth-like atom (such as Strontium or Ytterbium). Two-qubit entangling interactions are mediated with highly-excited Rydberg states \cite{saffman2010quantum}. 
In platforms that use a two-photon Rydberg transition, \texttt{CZ} and \texttt{Rz} gates are typically implemented via locally-addressing beams involving a blue-wavelength laser, which off-resonantly addresses an atomic transition between the computational $\ket{1}$ state and an intermediate state, and a global infrared laser, which bridges the excitation from the intermediate state to a Rydberg state. Simultaneous application of both the blue and infrared lasers on two nearby qubits achieves a \texttt{CZ} gate if the qubits are within a ``blockade radius'', i.e., if they are close enough that excitation to the Rydberg state in one atom shifts the Rydberg transition to be off-resonant in the other atom. Application of the blue laser alone induces an AC Stark shift, which shifts the $\ket{0}$ and $\ket{1}$ energy levels relative to each other without changing the state populations, thereby accomplishing a single-qubit \texttt{Rz} gate. Single-qubit \texttt{Rx} and \texttt{Ry} gates, which change the $\ket{0}$ and $\ket{1}$ state populations, require a different infrastructure, implemented via globally-addressing microwaves introduced with a microwave horn \cite{graham_rydberg_2019} or Raman laser system \cite{lukin_raman}.

\subsection{Related Work}\label{section: related work}

Numerous works exist on general quantum compilers or those tailored to superconducting systems, including \cite{anis_qiskit_2021,cirq_developers_2022_7465577,sivarajah_tket_2020,cowtan_qubit_2019,murali_noise-adaptive_2019,xu_quartz_2022,molavi_qubit_2022}. Ref.~\cite{schmid_computational_2024} provides an overview of current compiler capabilities for neutral atom quantum processors, which we summarize briefly here. Baker et al.~\cite{baker_exploiting_2021} proposed the first compiler to account for neutral atom hardware features and challenges, including long-distance interactions and atom loss. Patel et al.~\cite{patel_geyser_2022} expanded on this with compiler strategies to replace one- and two-qubit circuit blocks with native three-qubit gates, thereby reducing total pulse count. Tan et al.~\cite{tan_qubit_2022}, Schmid et al.~\cite{schmid_hybrid_2023}, and Wang et al.~\cite{wang_atomique_2024} incorporated physical atom movement \cite{bluvstein_quantum_2022} into routing. Li et al.~\cite{li_timing-aware_2023} developed scheduling techniques based on connectivity and parallelism constraints in neutral atoms. Yet none of these compilers support decompositions nor optimizations involving global gates, and modifying previous frameworks to incorporate this is nontrivial.

Some experimental work \cite{graham_multi-qubit_2022} has employed the straightforward decomposition $\texttt{R}(\theta,\phi)=\texttt{GR}(\frac{\pi}{2},\phi+\frac{\pi}{2})\cdot\texttt{Rz}(\theta)\cdot\texttt{GR}(-\frac{\pi}{2},\phi+\frac{\pi}{2})$ to achieve a universal set of local single-qubit gates. However, this is only valid for decomposing one gate at a time, rather than entire moments of parallel gates. When applied to a full circuit, the imposed serialization renders this method impractical. Other experimental work \cite{bluvstein_quantum_2022} uses a semi-global blue-wavelength laser combined with atom movement to achieve local \texttt{Rz} and \texttt{CZ} gates, with arbitrary single-qubit gates still requiring decomposition into local z-axis rotations and global x-axis and y-axis rotations. Neither approach can be easily incorporated into a systems-level framework without unreasonable gate counts, errors, and circuit durations.  

\begin{figure}
    \centering
    \begin{subfigure}[b]{0.95\columnwidth}
        \centering
        \includegraphics[width=\textwidth]{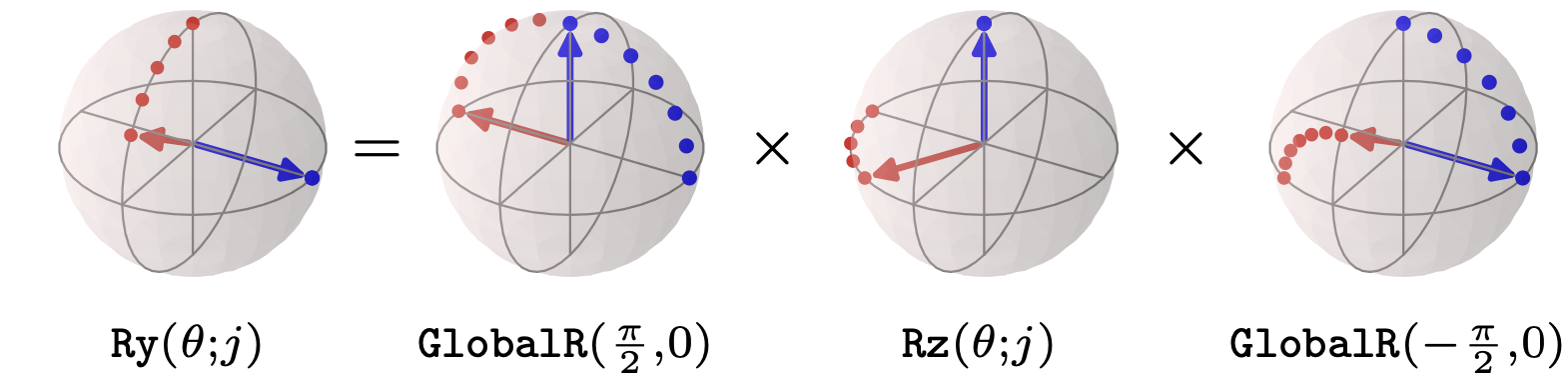}
        \caption{
        }
        \label{fig:decomposition_visualizations_axial}
     \end{subfigure}
     \\
    \begin{subfigure}[b]{0.95\columnwidth}
        \centering
        \includegraphics[width=\textwidth]{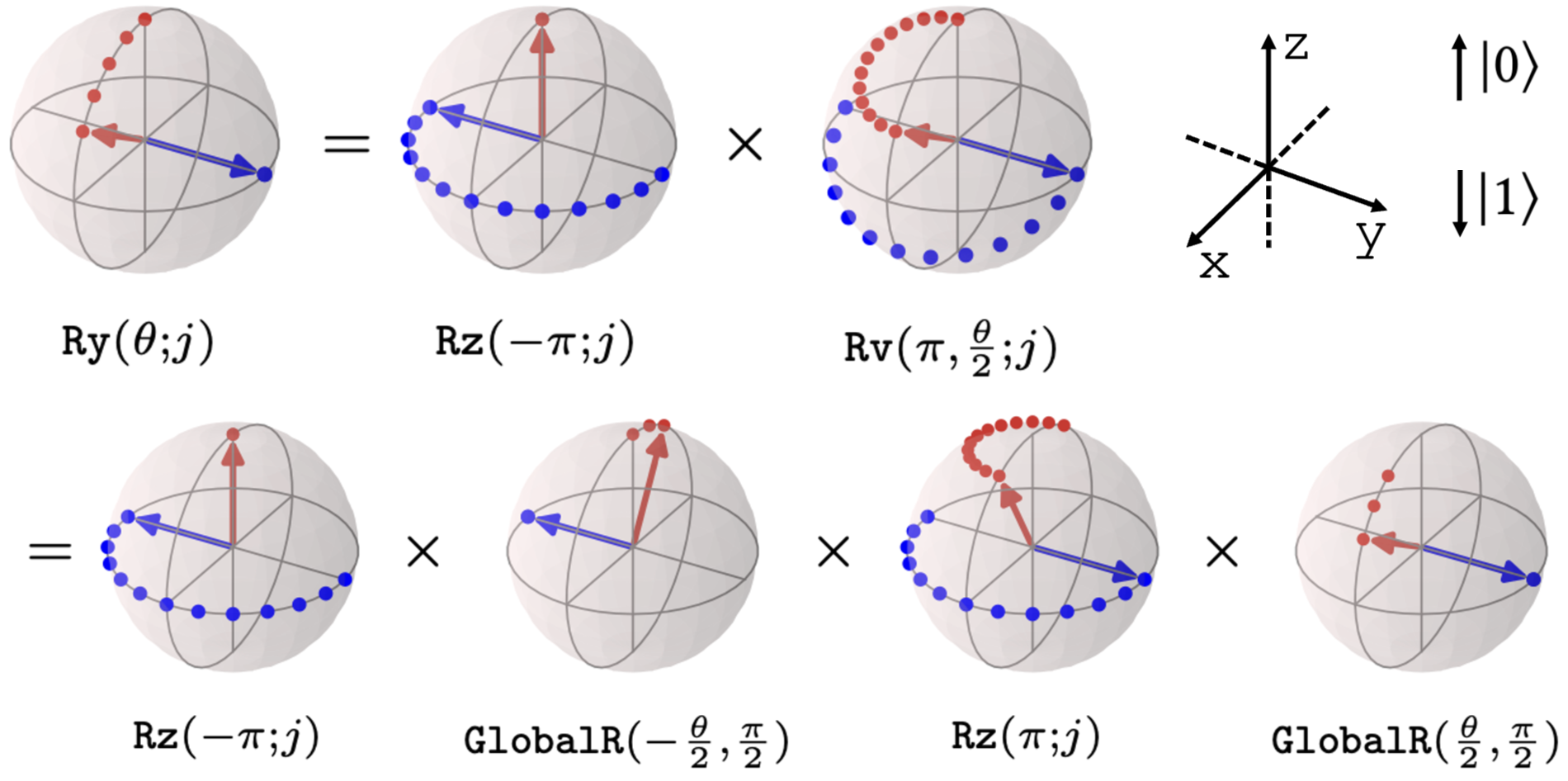}
        \caption{
        }
        \label{fig:decomposition_visualizations_transverse}
    \end{subfigure}
        \caption{
         Two decompositions of a local gate that rotates qubit $j$ about the \texttt{Y} axis by an angle $\theta$.
         Red and blue arrows track the trajectories of states initially pointing along the $+z$ and $+y$ axes of the Bloch sphere, respectively, corresponding to the initial states $\ket{0}$ and $\ket{0}+i\ket{1}$.
         (a) Axial decomposition of a local \texttt{Ry} gate, which uses global pulses to swap the \texttt{Y} and \texttt{Z} axes and imprints the rotation angle $\theta$ onto qubit $j$ with a local (axial) \texttt{Rz} gate. This involves a net global rotation of $\pi$. (b) Transverse decomposition of a local \texttt{Ry} gate, which uses global pulses to swap the $\texttt{V}_{\theta/2}$ and \texttt{Z} axes, thereby imprinting the rotation angle $\theta$ onto qubit $j$ with global (transverse) \texttt{GR} gates. This involves a net global rotation of $\lvert\theta\rvert\le\pi$.}
    \label{fig:decomposition_visualizations}
\end{figure}

\section{Compiler Pipeline Overview}\label{section: compiler pipeline overview}

We start by providing an overview of the steps in our compiler pipeline, shown in Figure \ref{fig: compiler}. The input into our compiler is a circuit in any arbitrary universal gate set. Hardware parameters, such as blockade radius and atom spacing, must also be specified. These are used to construct the device's \textit{connectivity graph}, where nodes represent atoms and the edge set contains all pairs of atoms that can interact via entangling gates. On neutral atoms, connectivity is determined by blockade radius, as defined in Sec.~\ref{section: hardware background}. 

The input circuit is then \textit{mapped} and \textit{routed}, i.e., \texttt{SWAP} gates or atom movement operations are inserted until all two-qubit gates act on connected pairs of atoms. This occurs at the beginning of the compiler pipeline so that the added gates are considered in the steps that follow. If the input circuit contains three-qubit gates such as Toffolis, these can be decomposed into one- and two-qubit gates before applying conventional routing techniques \cite{li_tackling_2019,cowtan_qubit_2019,sivarajah_tket_2020,anis_qiskit_2021}. Alternatively, three-qubit gates can be left in the circuit, and routing techniques can be adjusted to account for multi-qubit gates, as in \cite{baker_exploiting_2021}.

Next, the circuit is \textit{transpiled} to $\set{\texttt{U3},\texttt{CZ},\texttt{CCZ}}$, which we call the \texttt{IntermediateGateSet}, then \textit{pre-compiled} with optimizations to commute and cancel gates and merge adjacent single-qubit gates. For blocks of single-qubit and two-qubit gates, existing transpilation passes from \texttt{Qiskit} \cite{anis_qiskit_2021} or \texttt{Cirq} \cite{cirq_developers_2022_7465577} are sufficient for this, and these passes can be easily adapted if the circuit contains three-qubit gates.

The focus of this work is on \textit{scheduling} and \textit{decomposition} strategies that account for the use of global gates in neutral atom architectures. During scheduling, the circuit is split into \textit{Single-Qubit Gate Moments} (SQGMs) and \textit{Multi-Qubit Gate Moments} (MQGMs). Each SQGM is a set of \texttt{U3} gates that can be executed in parallel. Each MQGM contains a set of consecutive \texttt{CZ} and \texttt{CCZ} gates which, for now, may or may not be executable in parallel. If atom movement was used for routing, this should be included in the MQGMs as well. During decomposition, each SQGM is translated to \texttt{Rz} and \texttt{GR} gates, so that the final circuit is in the \texttt{NeutralAtomGateSet} $\set{\texttt{Rz},\texttt{GR},\texttt{CZ},\texttt{CCZ}}$. We describe this in Sections \ref{section: decomposition} and \ref{section: scheduling}.

Lastly, we enforce program-level and hardware-level parallelism constraints in MQGMs. Multiple \texttt{CZ} and \texttt{CCZ} gates are parallelizable if 1) the gates act on disjoint subsets of qubits, and 2) operands of \textit{different} gates are \textit{farther} than a blockade radius---e.g., \texttt{CZ}$\left(q_a,q_b\right)$ and \texttt{CZ}$\left(q_c,q_d\right)$ can execute simultaneously if none of $\set{\left(q_a,q_c\right), \left(q_a,q_d\right), \left(q_b,q_c\right), \left(q_b,q_d\right)}$ are in the connectivity graph's edge set. Each MQGM is split into sub-moments based on these constraints, where each sub-moment's gates can be executed simultaneously. Typically, atom movement is not performed in parallel with gates; while it is possible to move atoms during global gate execution, this may lead to increased atom loss and worsened gate fidelities. Different atom movement operations can occur in parallel with each other if they follow the constraints outlined in \cite{tan_qubit_2022}.

For the final output circuit, we calculate the circuit duration and fidelity as explained in Section \ref{section: metrics}.   

\section{Decomposition}\label{section: decomposition}

In this section, we detail how we translate a circuit into the \texttt{NeutralAtomGateSet}. We assume the circuit has previously been transpiled to the \texttt{IntermediateGateSet} and, importantly, scheduled into moments.  We describe the decomposition step first, despite it appearing \textit{after} scheduling in our compiler pipeline, because it motivates our approaches for scheduling. MQGMs are already in the native gate set, so decompositions only need to be applied to SQGMs.

\subsection{Motivation}

\begin{figure}
     \centering
     \begin{subfigure}[b]{0.48\textwidth}
         \centering
         \includegraphics[width=\textwidth]{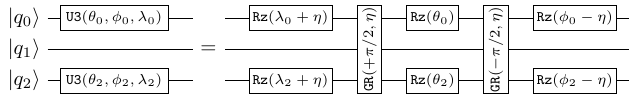}
         \caption{} 
         \label{fig:decomposition_circuits_axial}
     \end{subfigure}
     \hfill
     \begin{subfigure}[b]{0.48\textwidth}
         \centering
         \includegraphics[width=\textwidth]{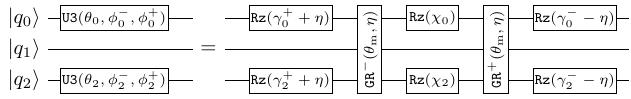}
         \caption{}
         \label{fig:decomposition_circuits_transverse}
     \end{subfigure}
        \caption{A moment of \texttt{U3} gates decomposed into the \texttt{NeutralAtomGateSet} using (a) Axial decomposition and (b) Transverse decomposition.
        Here $\texttt{GR}^\pm(\theta_{\mathrm{m}},\eta)=\texttt{GR}(\pm\theta_{\mathrm{m}}/2,\pi/2+\eta)$ and $\theta_{\mathrm{m}}=\pm\max_j\abs{\theta_j}$ for shorthand.
        The global rotations in both decompositions cancel on qubit $q_1$, which has no gate acting on it in the original moment.
        }
    \label{fig:decomposition_circuits}
\end{figure}

As discussed previously, executing quantum circuits on neutral atom platforms typically requires us to decompose into a native gate set involving global gates. Our Axial decomposition in \ref{section: axial decomposition} provides a simple and intuitive starting point for satisfying this compiler step, while maintaining the parallelism in the original circuit. However, it involves no other optimizations. The Transverse decomposition in \ref{section: transverse decomposition} is motivated by the fact that \texttt{GR} costs dominate gate execution times on many neutral atom architectures. For a given SQGM, we find the minimum \texttt{GR} rotation amount required to decompose into the \texttt{NeutralAtomGateSet}, with negligible to no increase in costs due to other gate types.

\subsection{Axial Decomposition}\label{section: axial decomposition}

We begin by decomposing any single-qubit \texttt{U3} gate into Euler-angle rotations, then further decomposing the \texttt{Ry} as 
\begin{align}
  \texttt{U3}(\theta, \phi, \lambda)
  &\simeq \texttt{Rz}(\phi)
  \, \texttt{Ry}(\theta)
  \, \texttt{Rz}(\lambda)\label{eqn:euler-angle}\\
  &= \texttt{Rz}(\phi)
  \, \texttt{Rx}\left(-\frac{\pi}{2}\right)
  \, \texttt{Rz}(\theta)
  \, \texttt{Rx}\left(\frac{\pi}{2}\right)
  \, \texttt{Rz}(\lambda),
  \label{eqn:ry_axial_decomp}
\end{align}
where $\simeq$ denotes equality up to a global phase; and \texttt{Rx}, \texttt{Ry}, and \texttt{Rz} are, respectively, single-qubit rotations about the $x$, $y$, and $z$ axes. The local \texttt{Rz} gate is already an element of the \texttt{NeutralAtomGateSet}. Within each SQGM, the local \texttt{Rx}$\left(-\frac{\pi}{2}\right)$ and \texttt{Rx}$\left(\frac{\pi}{2}\right)$ gates can be replaced by global \texttt{GR} gates, since the axis and angle of rotation are identical for all qubits. On qubits that were not acted upon by a \texttt{U3}, the \texttt{GR} gates cancel, preserving the original moment's overall operation. Columns of local \texttt{Rz} gates, in between and on either side of the two \texttt{GR} gates, are executed in parallel.
 
Together, the entire moment is decomposed into the form 

\begin{align}
  \prod_j \texttt{U3}_j(\theta_j,\phi_j,\lambda_j) =
  \left[\prod_j\texttt{Rz}_j(\phi_j)\right]\cdot
  \texttt{GR}\left(-\frac{\pi}{2},0\right)\notag\\
  \cdot\left[\prod_j\texttt{Rz}_j(\theta_j)\right]\cdot
  \texttt{GR}\left(\frac{\pi}{2},0\right)\cdot
  \left[\prod_j\texttt{Rz}_j(\lambda_j)\right],
  \label{eqn:axial_decomp}
\end{align}
where $\texttt{Rz}_j$ are the \texttt{Rz} gates addressing qubit $j$.
Intuitively, this decomposition uses a \texttt{GR} gate to move the $y$ axis to the $z$ axis, rotates qubits about the $z$ axis with local \texttt{Rz} gates, and then moves the $z$ axis back to the $y$ axis with another \texttt{GR} gate, thereby implementing site-specific \texttt{Ry} rotations. This decomposition is visualized on the Bloch sphere in Figure \ref{fig:decomposition_visualizations_axial} and shown in circuit form in Figure \ref{fig:decomposition_circuits_axial}.

\subsection{Transverse Decomposition}\label{section: transverse decomposition}

We again start by decomposing the \texttt{U3} gates according to the Euler-angle rotations in Eq.~\eqref{eqn:euler-angle}. The difference comes from our decomposition of the single-qubit \texttt{Ry} gate, which relies on decomposing the rotation into two reflections as

\begin{align}
  \texttt{Ry}(\theta)
  = \texttt{Rv}\left(\pi, \frac{\theta}{2}\right)\texttt{Rz}(-\pi)
  \label{eqn: Ry to Rv}
\end{align}
where $\texttt{Rv}(\xi,\omega)$ is a single-qubit rotation by the angle $\xi$ about the axis $\texttt{V}_\omega=\cos\omega\texttt{Z}+\sin\omega\texttt{X}$.
In turn, we can decompose
\begin{align}
  \texttt{Rv}(\xi, \omega)
  = \texttt{GR}\left(\omega, \frac{\pi}{2}\right)
  \texttt{Rz}(\xi)
  \texttt{GR}\left(-\omega, \frac{\pi}{2}\right).
  \label{eqn: Rv to GR}
\end{align}
Similarly to the Axial decomposition, Eq.~\eqref{eqn: Rv to GR} can be understood as using a \texttt{GR} gate to move the $\texttt{V}_\omega$ axis to the $z$ axis, rotating about the $z$ axis, and moving the $z$ axis back to $\texttt{V}_\omega$, as visualized in Figure \ref{fig:decomposition_visualizations_transverse}.

By combining Eq.~\eqref{eqn:euler-angle} and Eq.~\eqref{eqn: Ry to Rv}-\eqref{eqn: Rv to GR}, we can decompose a moment of multiple \texttt{U3} gates into the form
\begin{align}
  \prod_j\texttt{U3}_j(\theta_j, \phi_j^-, \phi_j^+) 
  = \prod_j
  \texttt{Rz}_j(\gamma_j^-)
  \texttt{Rv}_j\left(\chi_j, \frac{\theta_\tmax}{2}\right)
  \texttt{Rz}_j(\gamma_j^+)
  \label{eqn: transverse full moment}
\end{align}
where $\theta_\tmax=\pm\max_j\abs{\theta_j}$ (sign arbitrary), and the angles $\gamma_j^+$, $\gamma_j^-$, and $\chi_j$ are defined by the following expressions:
\begin{align}
  \gamma_j^\pm &= \left[\phi_j^\pm - \sigma_j (\alpha_j \pm \beta_j)\right] \mod 2\pi, \label{eqn: transverse gamma} \\ 
  \alpha_j &= \arctan(\cos(\theta_\tmax/2) \kappa_j), \label{eqn: transverse alpha}\\
  \beta_j &= \sign(\theta_j) \sign(\theta_\tmax) \times \frac{\pi}{2}, \label{eqn: transverse beta} \\
  \chi_j &= \left[\sigma_j \times 2\arctan(\kappa_j)\right] \mod 2\pi, \label{eqn: transverse chi} \\
  \kappa_j &= \sqrt{\frac{\sin(\theta_j/2)^2}{\sin(\theta_\tmax/2)^2-\sin(\theta_j/2)^2}}. \label{eqn: transverse kappa}
\end{align}
Here $\sign(x)=x/\abs{x}\in\set{+1,-1}$ if $x\ne 0$ and 0 otherwise; $\sigma_j\in\set{+1,-1}$ is an arbitrary sign; we define $\kappa_j = \infty$ if $\theta_j = \pm\theta_\tmax$, with $\arctan(\infty)=\pi/2$;
and ``mod $2\pi$'' is understood to mean that an angle is shifted to the interval $(-\pi,\pi]$.

The intuition behind the Transverse decomposition is that \texttt{Rv} changes the latitude (i.e., polar angle) of the states $\ket{0}$ and $\ket{1}$ on the Bloch sphere by an amount determined by $\chi_j$ and $\theta_\tmax$. When $\chi_j=\pi$, the polar angle changes by exactly $\abs{\theta_\tmax}$. When $\chi_j\neq\pi$, it changes by some $\abs\theta<\abs{\theta_\tmax}$. This explains both why we can use the same $\theta_\tmax$ angle to decompose all \texttt{U3} gates in the SQGM ($\chi_j$ can be tuned individually per qubit $j$ since it is the angle of a local \texttt{Rz} gate) and why $\theta_\tmax/2$ is the minimum rotation amount for each \texttt{GR} gate (less than this would not reach the latitudes necessary for every \texttt{U3} gate in the SQGM, regardless of the $\chi_j$ values). The \texttt{Rz} gates that sandwich the \texttt{Rv} change the longitude of a state on the Bloch sphere. Together, this allows us to implement any arbitrary single-qubit rotation (see Figures \ref{fig:decomposition_visualizations_transverse} and \ref{fig:decomposition_circuits_transverse}). We provide the derivations for Eq.~\ref{eqn: transverse full moment}-\ref{eqn: transverse kappa} in the Appendix.

%%%%%%%%%%%%%%%%%%%%%%%%%%%%%%%%%%%%%%%%%%%%%%%%%%
\subsection{Optimizing \texttt{Rz} Costs via Post-Processing}\label{section: post-processing}

Though the decompositions presented in Sec.~\ref{section: axial decomposition} and Sec.~\ref{section: transverse decomposition} are sufficient on their own, we can apply post-processing optimizations to additionally reduce \texttt{Rz} costs. One such simple compiler pass, which we implement in this work, eliminates the last layer of \texttt{Rz} gates in Eq.~\eqref{eqn:axial_decomp} and Eq.~\eqref{eqn: transverse full moment} by commuting them past adjacent \texttt{CZ} and \texttt{CCZ} gates and absorbing them into the next layer of \texttt{Rz} gates. In the case where $\theta_{max}=0$, the global gates are removed, and the surrounding \texttt{Rz} gates are similarly merged with the next layer. Additionally, the sign $\sigma_j$ in Eq.~\eqref{eqn: transverse chi}, the sign of $\theta_{max}$ in Eq.~\eqref{eqn: transverse full moment}, and $\eta$ in Fig.~\ref{fig:decomposition_circuits} can be chosen to achieve an auxiliary objective, such as further lowering \texttt{Rz} costs. %We choose not to focus on this last situation, since \texttt{Rz} gates have the lowest duration out of all native gates and thus will not have significant effect on the overall circuit costs.

\subsection{Time Complexity}
One major benefit of our decompositions is that, given a scheduling of gates, we require only linear time to find the solution with optimal \texttt{GR} rotation amount. For each \texttt{U3} gate, the final \texttt{Rz} and \texttt{GR} parameters are calculated by applying Eq.~\eqref{eqn:axial_decomp} (Axial) or Eq.~\eqref{eqn: Rv to GR}-\eqref{eqn: transverse kappa} (Transverse). For each \texttt{CZ} or \texttt{CCZ}, we append it to the circuit without changes. Both cases require constant time per gate. The total time complexity is therefore $\Theta(N_{gates})$, where $N_{gates}$ is the number of gates in the circuit. 

\section{Scheduling}\label{section: scheduling}

\begin{algorithm}[t!]
  \caption{Sifting through a circuit $C=(V,E)$.
    Here, $\texttt{topological\_order}(C)$ returns a sequence $(v_1,v_2,\cdots,v_{|V|})$ such that $j<k$ if $(v_j,v_k)\in E$.}
  \label{alg:sifting}
  \DontPrintSemicolon
  \SetKwComment{Comment}{\# }{}
  \SetKwProg{Fn}{def}{\string:}{}

  \SetKwInOut{Input}{Input}
  \SetKwInOut{Output}{Output}
  \SetKwFor{For}{for}{:}{}
  \SetKwIF{If}{ElseIf}{Else}{if}{:}{elif}{else:}{}
  \BlankLine
  \Input{circuit $C=(V,E)$ \\ indicator function $f:V\to\set{0,1}$}
  \Output{circuits $C_\passed$, $C_\caught$, $C_\remaining$}
  \BlankLine

  \SetKwFunction{Sift}{sift}
  \SetKwFunction{TopologicalOrder}{topological\_order}
  \SetKw{In}{in}
  \SetKwFunction{Qubits}{$\qubits$}

  \Fn{\Sift{$C,f$}}{
    $V_\passed, V_\caught, V_\remaining \leftarrow \set{}, \set{}, \set{}$ \\
    \For{operation $v$ \In \TopologicalOrder{$C$}}{
      \eIf{
        \Qubits{$v$} and \Qubits{$V_\caught \cup V_\remaining$} are disjoint
      }{
        \eIf{
          $f(v) = 0$
        }{
          add $v$ to $V_\passed$
        }{
          add $v$ to $V_\caught$
        }
      }{
        add $v$ to $V_\remaining$
      }
      \If{\Qubits{$V_\passed \cup V_\caught$} = \Qubits{$V$}}
      {\textbf{break}}
    }
    $C_\remaining \leftarrow$ subgraph of $C$ on $V_\remaining$\;
    \Return{$V_\passed$, $V_\caught$, $C_\remaining$}
  }
\end{algorithm}

We define a schedule as an assignment of gates to moments. Any schedule that splits the circuit into SQGMs and MQGMs, as defined in Sec.~\ref{section: compiler pipeline overview}, is a valid input into the decomposition step. But not all schedules that meet this requirement will perform equally once decomposed to the native gate set. Here, we discuss our proposed algorithms. 

\subsection{Motivation}\label{section: scheduling motivation}

The Transverse decomposition in Sec.~\ref{section: transverse decomposition} minimizes the \texttt{GR} costs \textit{for a given schedule}. By finding a better schedule\textemdash through approaches which specifically account for the equations in Sec.~\ref{section: decomposition}\textemdash we can further reduce these costs.

\begin{figure*}[t!]
    \centering

    \begin{subfigure}[t]
    {0.275\columnwidth}
        \centering
        \includegraphics[width=\textwidth]{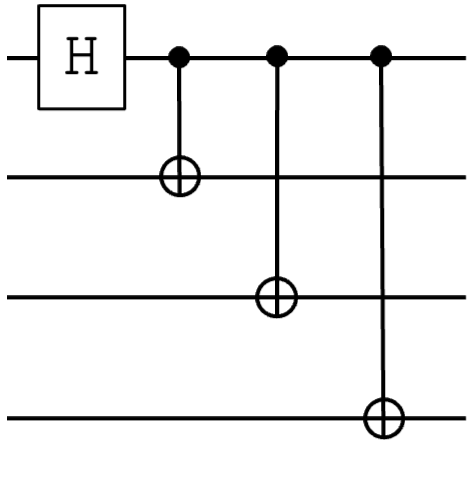}
        \caption{
        }
        \label{fig: ghz orig}
    \end{subfigure}\hspace{0.05\textwidth}
    ~
    \begin{subfigure}[t]{0.91\columnwidth}
        \centering
        \includegraphics[width=\textwidth]{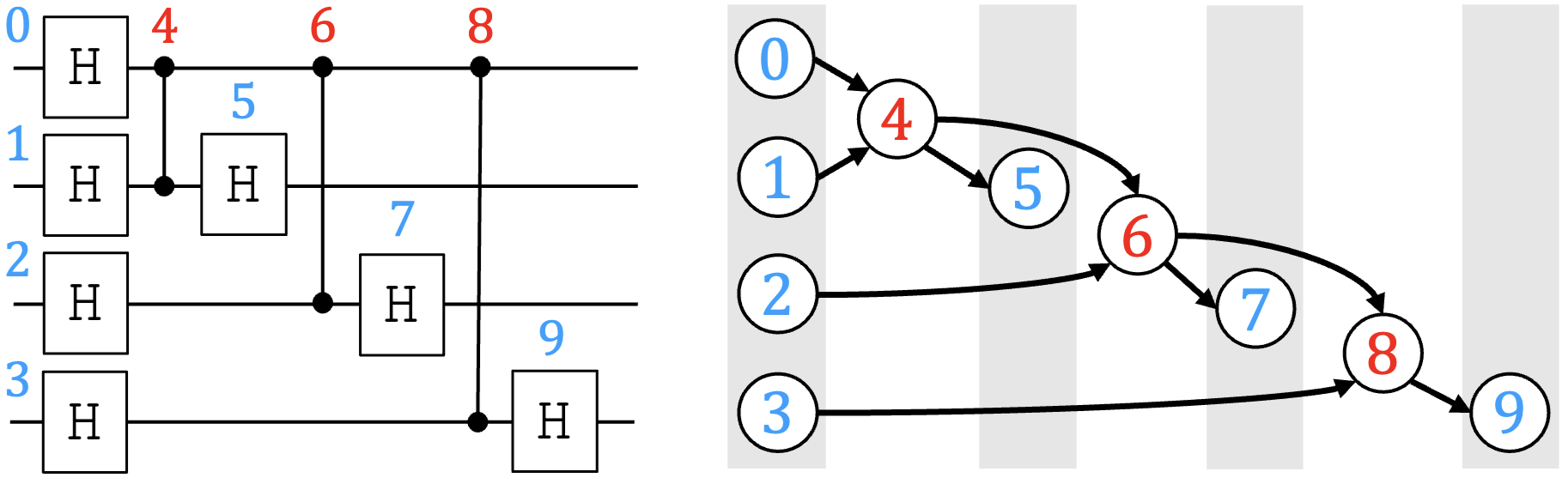}
        \caption{
        }
        \label{fig: ghz asap scheduling}
    \end{subfigure}\hspace{0.05\textwidth}
        ~ 
    \begin{subfigure}[t]{0.58\columnwidth}
        \centering
        \includegraphics[width=\textwidth]{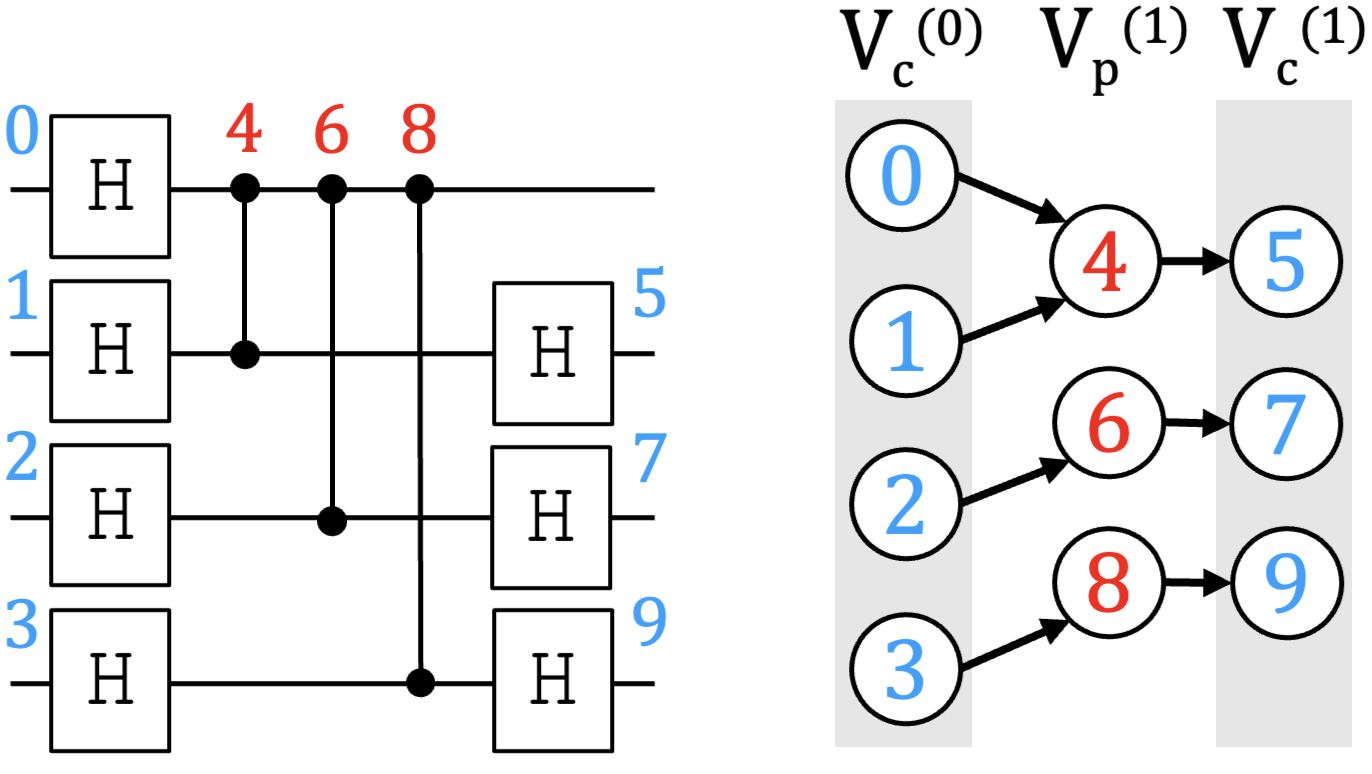}
        \caption{
        }
        \label{fig: ghz sfit scheduling}
     \end{subfigure}
    \caption{Motivation behind the Sifting scheduling algorithm. (a) A 4-qubit GHZ circuit before any compiler steps are applied. (b) The same circuit and its corresponding DAG when converted to the \texttt{IntermediateGateSet}, where \texttt{H}$=$\texttt{U3}$(\pi/2,0,\pi)$, and scheduled using an As-Soon-As-Possible approach. Grey rectangles show which gates are scheduled into the same SQGM. When decomposed into the \texttt{NeutralAtomGateSet}, this schedule will require 8 \texttt{GR} gates in total\textemdash two for each SQGM. (c) The schedule produced with Sifting. Now, only 4 \texttt{GR} gates are required in the final circuit. $V_p^{(i)}$ and $V_c^{(i)}$ are the $V_{passed}$ and $V_{caught}$ sets returned by the $i$th iteration of Sift, with $V_p^{(0)}=\emptyset$ here.}
    \label{fig: sifting motivation}
\end{figure*}

In Eq.~\eqref{eqn:axial_decomp} and Eq.~\eqref{eqn: transverse full moment}, exactly two \texttt{GR} gates are required to decompose each SQGM, regardless of how many \texttt{U3} gates are in the original moment. Thus, minimizing the number of SQGMs in the final schedule\textemdash in particular by maximizing parallelism of \texttt{U3} gates\textemdash will minimize the number of \texttt{GR} gates in the final circuit, as shown in Figure \ref{fig: sifting motivation}. This motivates our Sifting algorithm described in Sec.~\ref{section: sifting}.

\begin{figure*}[t!]
    \centering

    \begin{subfigure}[t]
    {0.37\columnwidth}
        \centering
        \includegraphics[width=\textwidth]{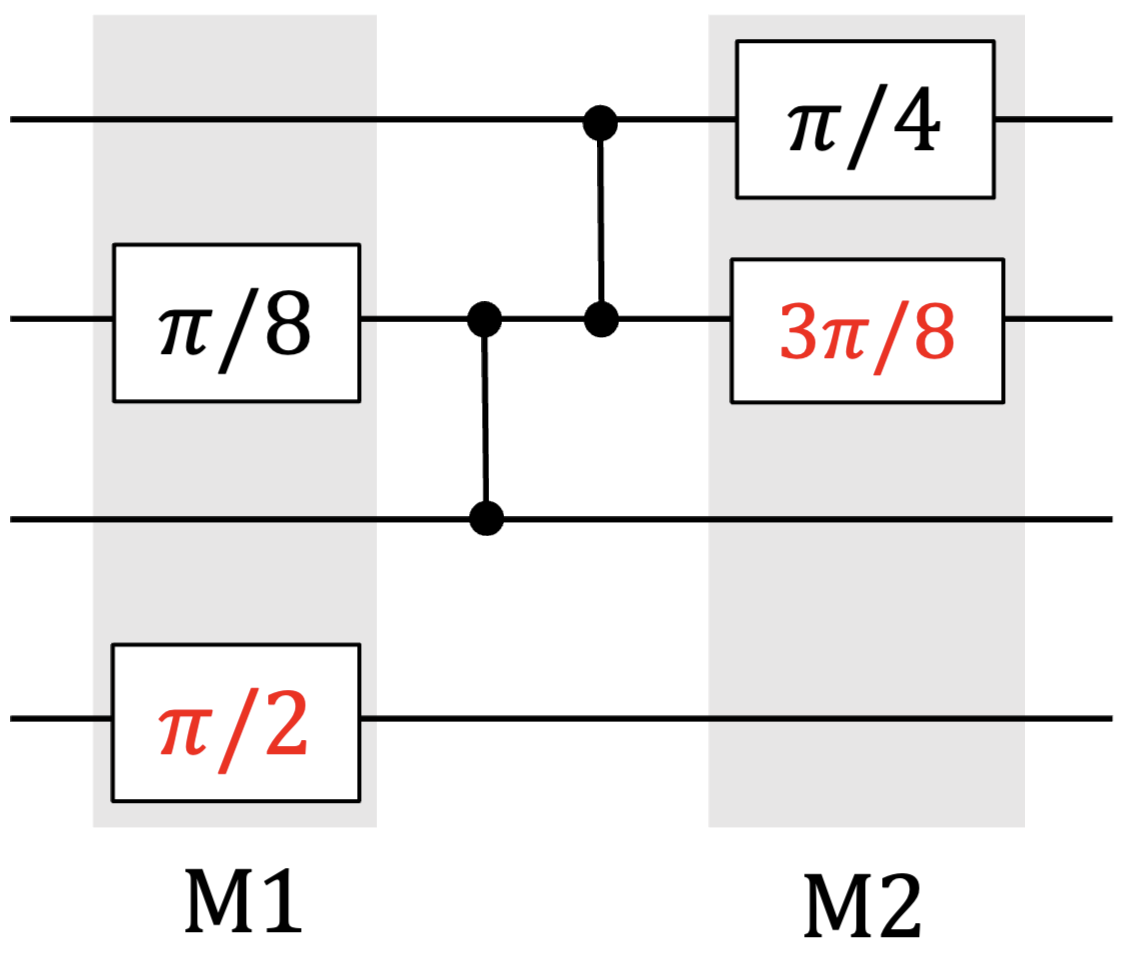}
        \caption{
        }
        \label{fig: dp motivation 1}
    \end{subfigure}\hspace{0.05\textwidth}
    ~
    \begin{subfigure}[t]{0.38\columnwidth}
        \centering
        \includegraphics[width=\textwidth]{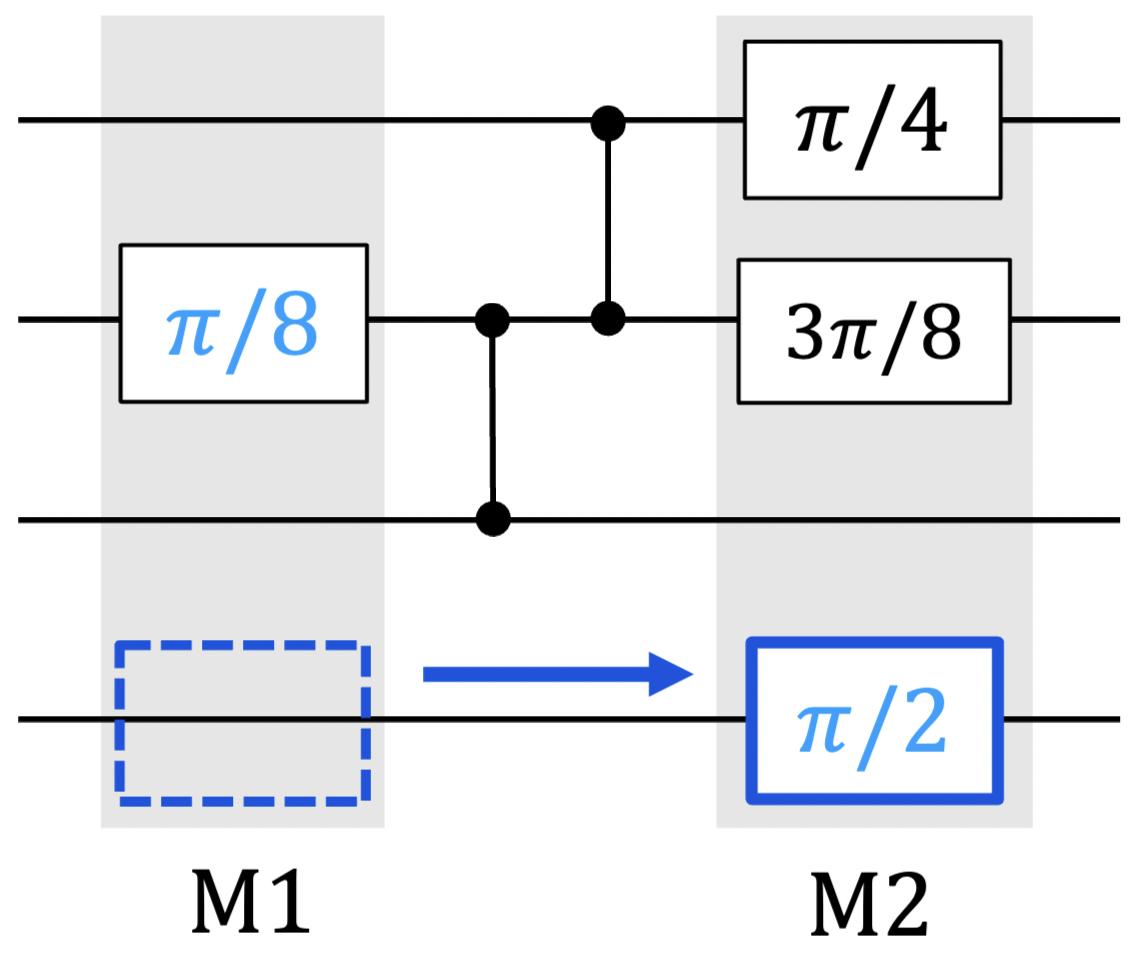}
        \caption{
        }
        \label{fig: dp motivation 2}
    \end{subfigure}\hspace{0.05\textwidth}
        ~ 
    \begin{subfigure}[t]{1.01\columnwidth}
        \centering
        \includegraphics[width=\textwidth]{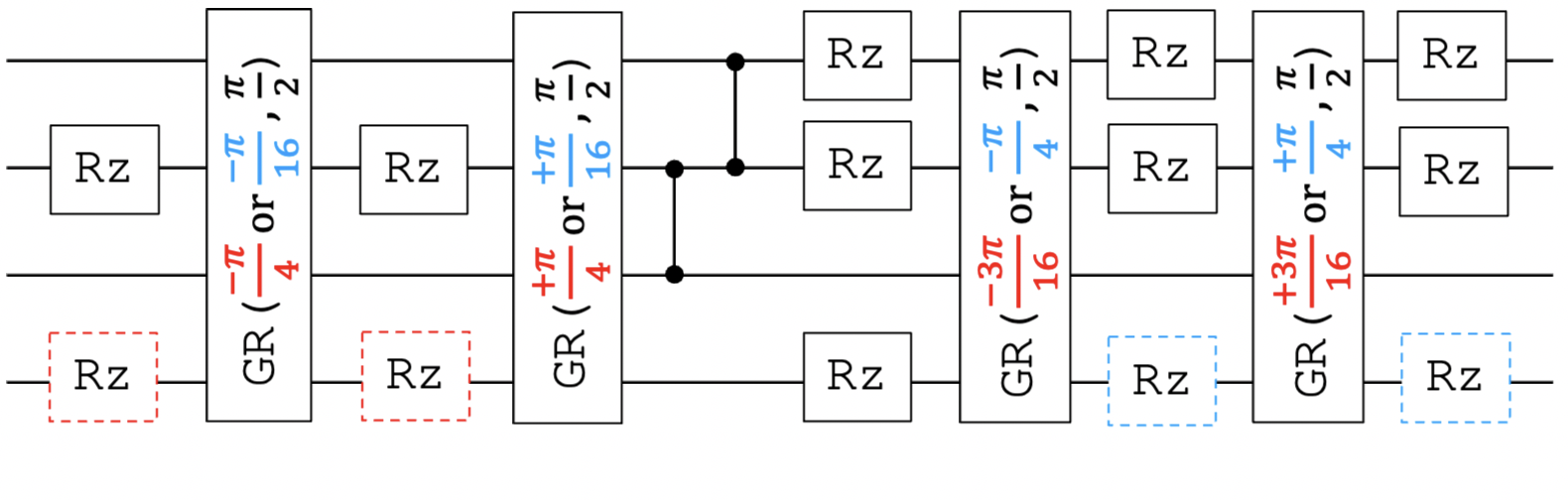}
        \caption{
        }
        \label{fig: dp motivation 3}
     \end{subfigure}
    \caption{Motivation behind the $\theta$-Opt algorithm. We show an example section of a circuit when (a) scheduled using Sifting and (b) scheduled using $\theta$-Opt, where the number shown in each single-qubit gate specifies the $\theta$ parameter. Both will require 4 \texttt{GR} gates when the Transverse decomposition is applied, shown in (c). However, the schedule in (a) results in a \texttt{GR} rotation amount of $\frac{\pi}{2}+\frac{3\pi}{8}=\frac{7\pi}{8}$, while the schedule in (b) results in $\frac{\pi}{8}+\frac{\pi}{2}=\frac{5\pi}{8}$. In (c), \texttt{GR} angles for (a) and (b) are given in red and blue, respectively. \texttt{Rz} gates outlined in red or blue dashed lines indicate those which appear in the decomposition of only (a) or only (b), respectively, while all other gates appear in both. Here, the last column of \texttt{Rz} gates from moment M1 is commuted past the \texttt{CZ} gates and combined with the first column of \texttt{Rz} gates from M2, as described in Sec.~\ref{section: post-processing}.}

    \label{fig: dp motivation}
\end{figure*}

We find for most circuits, however, we can achieve further improvement by minimizing not only the \textit{number} of \texttt{GR} gates, but also their \textit{rotation amount}. We use Sifting within our $\theta$-Opt algorithm in Sec.~\ref{section: dp algorithm}, which produces a schedule that\textemdash when combined with the Transverse decomposition\textemdash minimizes the total \texttt{GR} \textit{rotation amount} in the final circuit. Our algorithm is based on the observation that, according to Eq.~\eqref{eqn: Rv to GR}-\eqref{eqn: transverse full moment}, the \texttt{GR} rotation angle required to decompose a given SQGM equals the maximum Euler angle $\theta$ parameter of all \texttt{U3} gates in that moment. This is illustrated in Figure \ref{fig: dp motivation}.

\subsection{Sifting Through a Circuit} \label{section: sifting}

The Sifting algorithm produces a schedule such that: 1) operations within different specified categories are not placed in the same moment, and 2) parallelism of operations in one such category is maximized. These categories are designated by an indicator function $f:V_{dag}\to\set{0,1}$. For the purposes of this work, we define $f(v)=1$ if $v$ is a \texttt{U3} gate and $f(v)=0$ otherwise, ensuring the resulting schedule is a collection of SQGMs and MQGMs with \texttt{U3} parallelism maximized. However, we leave Alg.~\ref{alg:sifting} agnostic to gate type, as we believe it may have broader uses in the development of compilers that emphasize parallelization capabilities.

The input circuit $C=(V_{dag},E_{dag})$ is represented in the form of a directed acyclic graph (DAG), which captures the circuit's dependencies. Each node $v\in V_{dag}$ corresponds to a gate in the circuit, and an edge $(u,w)\in E_{dag}$ indicates that $u$ must complete prior to executing $w$. A call to Sift, given by Alg.~\ref{alg:sifting}, yields three subcircuits of $C$:
\begin{enumerate}
  \item $C_\passed$, with $f(v)=0$ for all gates $v$ in $V_\passed$.
  \item $C_\caught$, with $f(v)=1$ for all gates $v$ in $V_\caught$.
  \item $C_\remaining$, with every gate in $V_\remaining$ a successor of some gate in $V_\caught$.
\end{enumerate}
Concatenating $C_\passed$, $C_\caught$, and $C_\remaining$ recovers the original circuit $C$. To obtain the full schedule, $V_\passed$ then $V_\caught$ are appended as moments in the schedule (importantly in this order to respect dependencies), and Sift is called repeatedly on $C_\remaining$ until $C_{remaining}$ is empty.

\subsection{$\theta$-Opt Scheduling Algorithm} \label{section: dp algorithm}

We use a dynamic programming algorithm to find the schedule $S$ that minimizes the objective

\begin{align}
   \mathrm{cost} &=\sum_{M_j\in S}\begin{dcases}
    \mathrm{max}(\{g.\theta | g\in M_j\}),&\mathrm{if\:}M_j \mathrm{\:is\:a\:SQGM}\\
    0,&\mathrm{if\:}M_j\mathrm{\:is\:a\:MQGM}
    \end{dcases}
    \label{eqn: dp objective}
\end{align}
which equals the total \texttt{GR} rotation amount in the final circuit when using the Transverse decomposition. If desired, Eq.~\ref{eqn: dp objective} can be adjusted to incorporate the costs of other gate types.

At each dynamic programming call on a circuit $C$, defined by its gate set $V$ and dependencies $E_{dag}$, we set the next MQGM to be the subset $V_{passed}$ returned when calling Sift on $C$. We then ``try'' all potential next SQGMs and choose the one that results in the lowest overall cost. The cost of using a given $M_k$ as our next SQGM is the sum of two components: first, the cost incurred by $M_k$ itself, as defined in Eq.~\eqref{eqn: dp objective}; and second, the cost from scheduling the rest of the circuit, determined by the dynamic programming call on the subcircuit $C'$, consisting of the set of remaining (unscheduled) gates $V'=V-V_{passed}-M_k$. This is captured by the recurrence relation below:
\begin{align}
    F(C) = \begin{dcases}
        0, &V=\emptyset \\
        \mathrm{min}\{\mathrm{max}\left(\{g.\theta | g\in M_k\}\right)+F(C'),\forall k\}, &\mathrm{else}
    \end{dcases}
    \label{eqn: dp recurrence relation}
\end{align} 

Eq.~\ref{eqn: dp recurrence relation} gives us almost all we need to implement the scheduling algorithm. What remains is to determine the set of moments $\set{M_k}$ to consider. We observe that the $V_{caught}$ returned by Sift contains every \texttt{U3} that \textit{could} appear in the next SQGM, i.e., every \texttt{U3} whose dependencies are already scheduled. Thus, any subset of $V_{caught}$ is technically a valid next SQGM. A naive solution would be to try all $M_k\subseteq V_{caught}$. With a few observations, however, we can significantly reduce the algorithm's time complexity. We describe this below. 

\subsection{Eliminating Unnecessary Calls}\label{section: eliminating calls} 

We can prove in advance that the majority of subsets of $V_{caught}$ can be ignored, exponentially reducing the number of dynamic programming calls. To better illustrate this, we introduce new notation. $V_p^i$, $V_c^i$, $V_p^{i+1}$, and $V_c^{i+1}$ are the $V_{passed}$ and $V_{caught}$ moments returned by the current and subsequent iterations of Sift (called on subcircuits with gate sets $V$ and $V'=V-V_p^i-V_c^i$, respectively), and $V_{rem}$ includes all remaining gates. $M_k$, as above, is the subset of gates in $V_c^i$ that are kept in their original SQGM, while $m_k=V_c^i-M_k$ contains the gates in $V_c^i$ that are moved to the following SQGM (i.e., combined with gates in $V_c^{i+1}$), such that $M_k\cup m_k=V_c^i$ and $M_k\cap m_k=\emptyset$. To respect dependencies, when $m_k$ is pushed to a later moment, we must also push back every gate in $V_p^{i+1}$ that depends on any gate in $m_k$. Let $V_p^{i+1,\square}\subset V_p^{i+1}$ consist of those gates that depend on $m_k$, and let $V_p^{i+1,\star}=V_p^{i+1}-V_p^{i+1,\square}$ be the subset of gates that don't depend on $m_k$ and thus are kept in their original MQGM. Similarly, let $V_c^{i+1,\square}$ and $V_c^{i+1,\star}$ be the disjoint subsets of $V_c^{i+1}$ that do and don't depend on $V_p^{i+1,\square}$, respectively, where gates in $V_c^{i+1,\square}$ are pushed back to respect dependencies. The updated groupings of gates (denoted here with $\nu$) are therefore $\nu_p^i=V_p^i$, $\nu_c^i=M_k$, $\nu_p^{i+1}=V_p^{i+1,\star}$, $\nu_c^{i+1}=m_k\cup V_c^{i+1,\star}$, and $\nu_{rem}=V_p^{i+1,\square}\cup V_c^{i+1,\square}\cup V_{rem}$.

%We can prove in advance that the majority of subsets of $V_{caught}$ can be ignored, exponentially reducing the number of dynamic programming calls necessary. To better illustrate this, we introduce new notation. $V_p^i$, $V_c^i$, $V_p^{i+1}$, and $V_c^{i+1}$ are the $V_{passed}$ and $V_{caught}$ moments returned by the current and subsequent iterations of Sift (called on subcircuits with gate sets $V$ and $V'=V-V_p^i-V_c^i$, respectively), and $V_{rem}$ includes all remaining gates. $M_k$, as above, is the subset of gates in $V_c^i$ that are kept in their original SQGM, while $m_k$ are those moved to the following SQGM, so that $M_k\cup m_k=V_c^i$ and $M_k\cap m_k=\emptyset$. To respect dependencies, when $m_k$ is pushed to a later moment, we must also push back all gates in $V_p^{i+1}$ that depend on $m_k$. We call this subset $V_p^{i+1,\square}$, and $V_p^{i+1,\star}=V_p^{i+1}-V_p^{i+1,\square}$ are the gates that don't depend on $m_k$. Similarly, all gates in $V_c^{i+1}$ that depend on $V_p^{i+1,\square}$ must also be pushed back; $V_c^{i+1,\square},V_c^{i+1,\star}\subseteq V_c^{i+1}$ are the disjoint subsets that do and don't depend on $V_p^{i+1,\square}$. The new groupings of gates are now $v_p^i=V_p^i$, $v_c^i=M_k$, $v_p^{i+1}=V_p^{i+1,\star}$, $v_c^{i+1}=m_k\cup V_c^{i+1,\star}$, and $v_{rem}=V_p^{i+1,\square}\cup V_c^{i+1,\square}\cup V_{rem}$.

Using this notation, we specify conditions that must be met for a given $M_k\subseteq V_c^i$ to be considered in Eq.~\ref{eqn: dp recurrence relation}. If any of these fails, $M_k$ is guaranteed to give a worse overall cost than another $M_{k}'$ which is considered, and so $M_k$ is ignored. 

\underline{\textit{Condition 1:}} $\forall g\in M_k,\forall h\in m_k, g.\theta>h.\theta$. I.e., if $\theta_{max}$ is the maximum $\theta$ value of gates in $M_k$, then any $h\in V_c^i$ with $h.\theta\leq\theta_{max}$ must also be placed in $M_k$. Doing so never increases the cost of the current moment $\nu_c^i=M_k$, since cost is determined solely by $\theta_{max}$, nor will it restrict scheduling later in the circuit. Placing $h$ in $m_k$, however, could impose additional dependencies with later moments, thus limiting future choices (note this is true in general, but if $h.\theta>\theta_{max}$, the benefit of moving $h$ to $m_k$ may outweigh the cost of extra dependencies, which never occurs when $h.\theta\leq\theta_{max}$). 

\underline{\textit{Condition 2:}} $V_p^{i+1,\star}\neq \emptyset$, or equivalently, $V_p^{i+1,\square}\neq V_p^{i+1}$. This condition fails when every gate in $V_p^{i+1}$ depends on at least one gate in $m_k$, i.e., $\forall g\in V_p^{i+1}, \exists h\in m_k$ s.t. $(h,g)\in E_{dag}$. When this happens, $\nu_p^{i+1}$ is an empty MQGM so $\nu_c^{i+1}$ and $\nu_c^i$ can be combined into the same SQGM, resulting in a case equivalent to when $M_k=V_c^i$ and $m_k=\emptyset$. 

\underline{\textit{Condition 3:}} $V_c^{i+1,\star}\neq\emptyset$, or equivalently, $V_c^{i+1,\square}\neq V_c^{i+1}$. This condition fails when every gate in $V_c^{i+1}$ depends on at least one gate in $V_p^{i+1,\square}$, i.e., $\forall g\in V_c^{i+1}, \exists h\in V_p^{i+1,\square}$ s.t. $(h,g)\in E_{dag}$. When that happens, $m_k$ becomes its own SQGM, instead of being combined with a subset of gates in $V_c^{i+1}$. This is guaranteed to never give a lower total cost than the case where $M_k=V_c^i$ and $m_k=\emptyset$.

\subsection{Graph-Based Representation}

The recursive calls of the dynamic programming algorithm can be converted into a graph-based representation, while still relying on memoization. The benefit of this is twofold. First, while the recursive version allows us to easily calculate the optimal \textit{cost}, it is difficult to keep track of the actual \textit{schedule} without a graph. Second, we can implement tree pruning to reduce the compile time without compromising optimality.

Nodes in the graph represent recursive calls, each with an associated MQGM and SQGM, with directed edges from parent to child calls. Additionally, each node $v$ is assigned a value, equivalent to the cost of the (partial) schedule from source node to $v$. We add nodes in depth-first order, keeping track of the best ``end'' node that corresponds to a complete schedule. When a new node is added to the graph, its value is compared to that of the best end node found so far; if greater, we stop adding nodes along that branch. The final optimal schedule is composed by tracing from the best end node back to the source node---requiring time linear in the circuit depth---then reversing the order of the moments.  

The above is equivalent to naive recursion without memoization; by itself, it would lead to recomputing subproblems that have already been solved. To avoid recomputing, we make the following adjustments. As defined by Eq.~\ref{eqn: dp recurrence relation}, subproblems are specified by the remaining subset of unscheduled gates. Let $(x,y)$ be a directed edge, and let $G'$ be the subset of gates that were remaining when $y$ was added to the graph. Assume at a later point, while exploring a different potential schedule, we add a node $z$ with the same subset $G'$ of remaining gates. Because of the depth-first ordering, we are guaranteed to have already solved the subproblem for $G'$---given by the descendants of node $y$---and do not need to continue exploring along the branch containing $z$. Note, however, we may have just found a better solution for the first part of the circuit; the paths from source to $y$ and from source to $z$ give two possible schedules (which we will call $S_y$ and $S_z$) for the same subset of gates $G-G'$, where $G$ is the set of all gates in the original circuit. If the cost of schedule $S_z$ is less than the cost of $S_y$, we replace the edge $(x,y)$ with the edge $(x,z)$. We then reduce the node value of all descendants of $y$ by $(S_y-S_z)$.

%\subsection{Circuit blocking.} \label{section: circuit blocking}

%For larger circuits, we adapt circuit blocking \cite{wu_qgo_2022} to better work with our scheduling algorithm. The benefit of circuit blocking is that time complexity depends on block depth $k$ instead of the full circuit depth. Standard techniques, however, prevent us from optimizing across the boundaries of adjacent but separate blocks, which negatively impacts the effectiveness of our $\theta$-Opt algorithm. To fix this, we allow blocks to overlap by some defined ``boundary depth'' $l$. Specifically, the first $k$ moments from the Sifted schedule are run through $\theta$-Opt, and the first $k-l$ of these optimized moments are appended to the final schedule. The remaining $l$ optimized moments from that block are combined with the next $k-l$ unoptimized moments from the Sifted version, forming the next block to optimize via $\theta$-Opt. For simplicity, we restrict $k$ and $l$ to be even numbers, so that each block begins with an MQGM, as this works better with the structure of $\theta$-Opt. By correctly tuning the parameters $k$ and $l$, we achieve better scalability, with at most a negligible loss in optimality of the final solution.   
   
\subsection{Time Complexity}

Scheduling via Sifting, as described in Sec.~\ref{section: sifting}, requires $\mathcal{O}(d\cdot N_{gates})$ time, where $d$ and $N_{gates}$ are circuit depth and number of gates, respectively. Alg.~\ref{alg:sifting} requires $\mathcal{O}(N_{gates})$ time per call, with $\Theta(d)$ calls required to schedule the whole circuit.

When scheduling via $\theta$-Opt, the recursive case of each dynamic programming call---or equivalently, each node added if using the graph-based approach---requires at most $\mathcal{O}(n^3)$ time, where $n$ is the number of qubits. The majority of this time comes from the implementation of Sec.~\ref{section: eliminating calls} (which slightly increases time per call, but greatly reduces the total number of calls). In many cases, however, time per call is closer to the lower bound of $\Omega(n\log{}n)$. The total number of dynamic programming calls is best case $\Omega(d\cdot n\log{}n)$ and worst case $\mathcal{O}(n^d)$---though importantly, in practice, this absolute worst case complexity is often not required. For very large circuit sizes, circuit blocking techniques \cite{wu_qgo_2022} can easily be adapted to work with our scheduling algorithm.

%, especially if applying Sec.~\ref{section: eliminating calls}-\ref{section: circuit blocking}. We show in Sec.~\ref{section: results} that our algorithm scales well enough to compile circuits of width \textcolor{red}{[fill in]} and depth \textcolor{red}{[fill in]}.

\section{Metrics}\label{section: metrics}

\begin{figure*}
    \centering
    \includegraphics[scale=0.47]{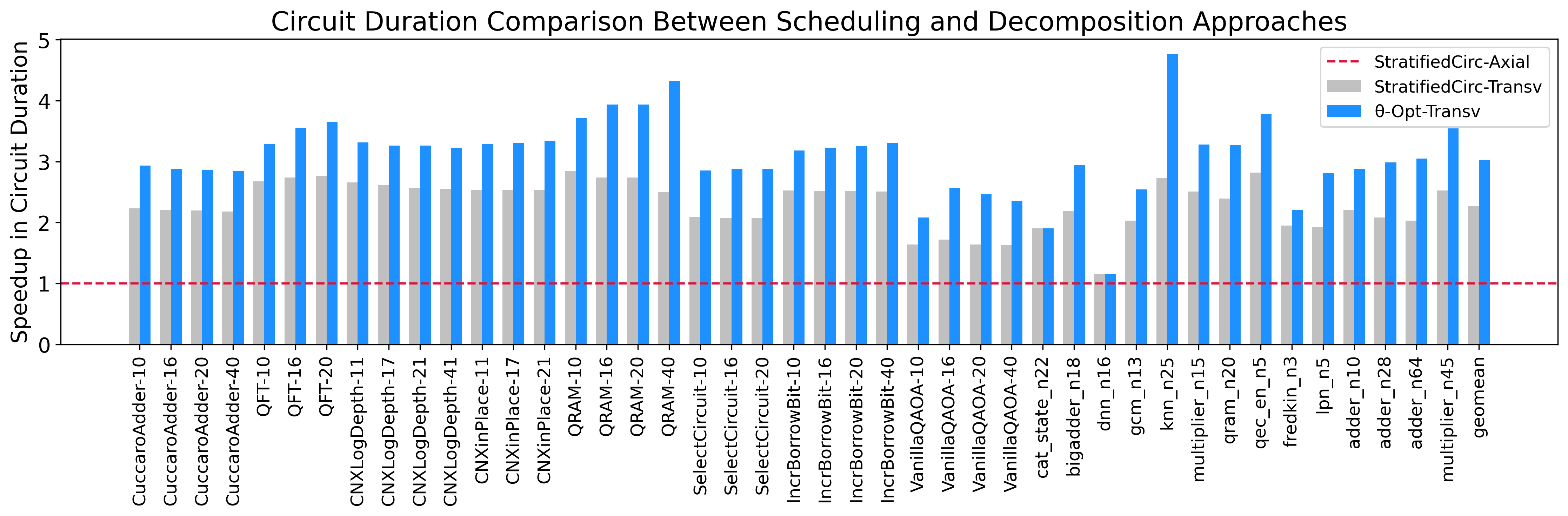}
    \caption{Speedups in circuit duration for (1) \texttt{stratified\_circuit} scheduling + Transverse decomposition and (2) $\theta$-Opt scheduling + Transverse decomposition. Both are relative to the \texttt{stratified\_circuit} + Axial combination. The $\theta$-Opt + Axial combination is the same as \texttt{stratified\_circuit} + Axial, since $\theta$-Opt scheduling only impacts the Transverse decomposition's \texttt{GR} rotation angles; Axial uses the same \texttt{GR} angles regardless of how we schedule.}
    \label{fig: results circuit duration speedup}
\end{figure*}

\subsection{Circuit Duration}\label{section: circuit duration calculations}

Once all compiler steps outlined in Section \ref{section: compiler pipeline overview} have been executed, the circuit structure is naturally split into moments of a) parallel \texttt{Rz} gates, b) parallel \texttt{CZ} and \texttt{CCZ} gates, and c) single \texttt{GR} gates. The duration of each moment is equal to the duration of the longest gate within that moment, and the total circuit duration is calculated by summing over the durations of all moments. For \texttt{Rz} and \texttt{GR} gates, we specify time required to execute a rotation of $\pi$ (determined by the Rabi frequencies), and duration of each gate scales linearly with rotation angle. 

\subsection{Fidelity}\label{section: fidelity calculations}

We account for the impact of both gate errors and idle errors on circuit fidelity. Error rates are chosen to reflect values from recent experimental work \cite{evered_high-fidelity_2023,graham_multi-qubit_2022,bluvstein_quantum_2022}. Microwave-based control techniques are typically limited by two error sources: Rabi frequency inhomogeneity across the array, resulting in over- or under-rotation of some qubits, and; stochastic Doppler shifts from motion of the finite-temperature atoms, adding a small detuning to the microwave drive. Both of these effects introduce infidelity which is second-order in the duration of the pulse, so the fidelity of a microwave $\texttt{GR}(\theta,\phi)$ gate is modeled as $\mathcal{F}=1-C_{GR}\cdot\theta^2$ for some constant $C_{GR}$. Control techniques using far off-resonant laser light, such as Stark shift lasers for implementing $\texttt{Rz}(\lambda)$ gates, are instead limited by photon scattering from the off-resonant state. This constant rate of scattering causes an exponential decay of fidelity, so for the short pulses used to implement $\texttt{Rz}(\lambda)$, the fidelity is approximately $\mathcal{F}=1-C_{Rz}\cdot\lambda$ for some $C_{Rz}$. Entangling gates such as $\texttt{CZ}$ and $\texttt{CCZ}$ face many sources of error, but since they are implemented using a fixed pulse sequence, we simply use the experimentally reported fidelity per gate.

The circuit's idle error due to dephasing is modeled as $1-e^{t/\mathrm{T_2^*}}$, where $t$ is the circuit's total duration, and $T_2^*$ is an experimentally-determined dephasing rate. For neutral atoms, amplitude damping is negligible relative to dephasing and other error sources and can be ignored in our calculations. 

Because cost of full simulation scales exponentially with the quantum system's size, and because this would be unrealistic for many of the benchmark sizes that we test, we estimate the circuit fidelity as the product of its idle error and all individual gate errors. This gives an upper bound on the fidelity.

\section{Evaluation}\label{section: evaluation}

We evaluate our compiler on a variety of benchmarks \cite{qcb,tomesh_supermarq_2022,li_qasmbench_2022}, chosen to reflect relevant quantum computing programs. These include subroutines, arithmetic circuits, and application circuits, among others. Additionally, to demonstrate scalability, we select a wide range of circuit sizes. Each benchmark circuit is run through the compiler pipeline described in Sec.~\ref{section: compiler pipeline overview}. Unless otherwise specified, we use Qiskit's \cite{anis_qiskit_2021} implementation of the routing method in \cite{li_tackling_2019}, with a blockade radius to atom spacing ratio of 3:1 when constructing the connectivity graph. For each benchmark, we compare our $\theta$-Opt algorithm with Cirq's \cite{cirq_developers_2022_7465577} \texttt{stratified\_circuit}, and our Transverse decomposition to our Axial decomposition. These comparisons demonstrate the benefit gained by specifically minimizing \texttt{GR} rotation amounts. We additionally compare to the closest prior existing work \cite{graham_multi-qubit_2022}, which was described briefly in Sec.~\ref{section: related work}.

When calculating circuit duration and fidelity as explained in Sec.~\ref{section: metrics}, we assume \texttt{CZ} and \texttt{CCZ} gate durations of 270 ns and 390 ns, respectively, with fidelities of 99.5\% and 97.9\% \cite{evered_high-fidelity_2023}. Rabi frequencies for \texttt{Rz} and \texttt{GR} gates are set to 3 MHz and 76.5 kHz, respectively, and $\mathrm{T_2^*}$ is set to 4 ms. With these parameters, experiments have achieved pulse fidelities of about 99.5\% for \texttt{Rz}$(\pi)$ and 99.8\% for \texttt{GR}$(7\pi/4,\phi)$, where the given rotation angles are an average during some benchmarking process \cite{bluvstein_quantum_2022,graham_multi-qubit_2022,jiang_sensitivity_2023}. Matching these values, we model the angle-dependent gate fidelities as $\mathcal{F}=1-.005 (\lambda/\pi)$ for \texttt{Rz}$(\lambda)$, and $\mathcal{F}=1-.002 (4\theta/7\pi)^2$ for \texttt{GR}$(\theta,\phi)$. 

\section{Results \& Discussion}\label{section: results}

\begin{figure}
\includegraphics[scale=0.34
]{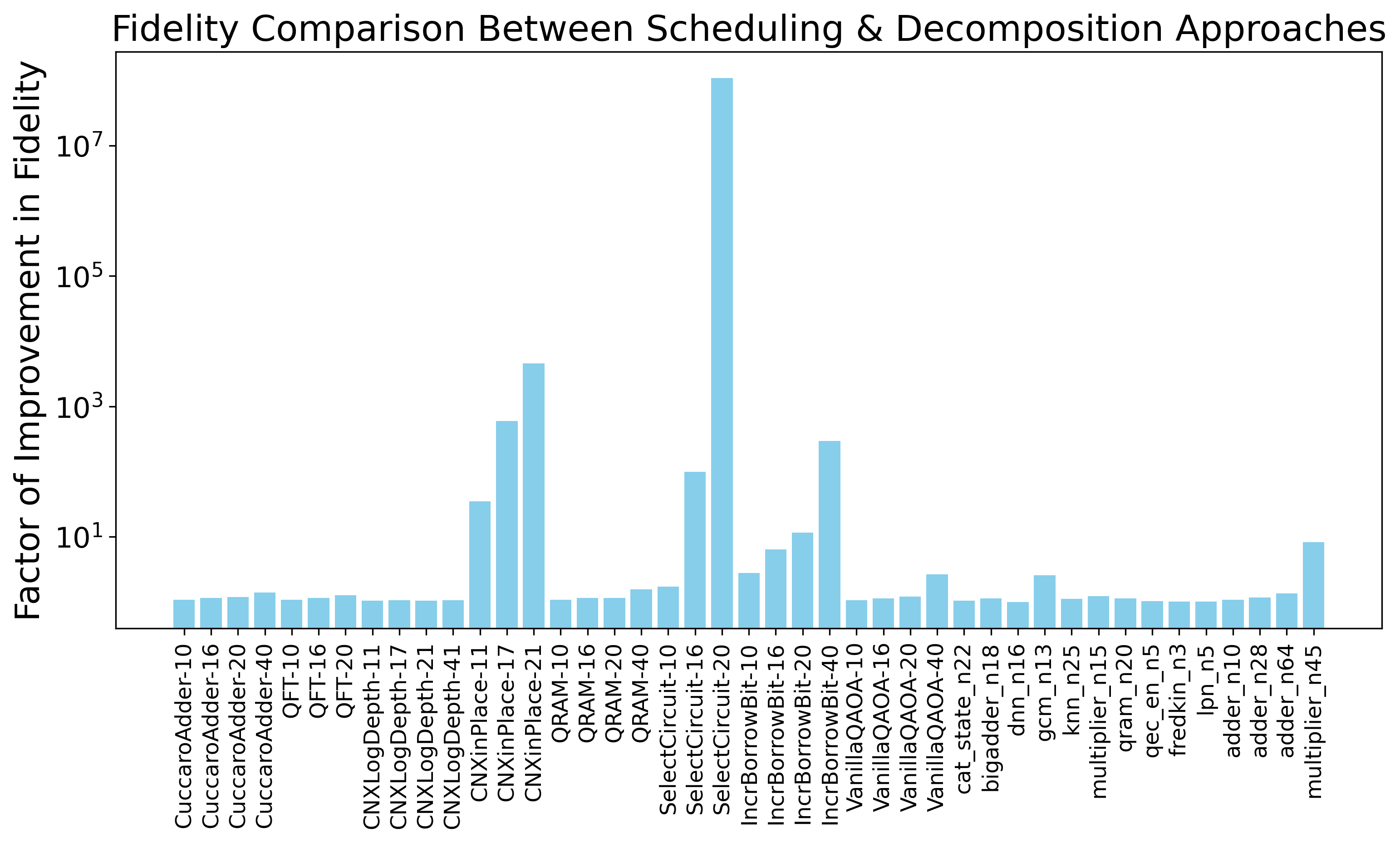}
    \caption{Factor of improvement in circuit fidelity that is achieved when combining the $\theta$-Opt scheduling with the Transverse decomposition, relative to the combination of \texttt{stratified\_circuit} with the Axial decomposition.}
    \label{fig: results fidelities}
\end{figure}

\subsection{Speedup in Circuit Duration}\label{section: results circuit duration}

We first compare our final compiler pipeline to \cite{graham_multi-qubit_2022}, achieving up to 53.8x speedup, with a geometric mean of 22.7x. This is due to our compiler's ability to maintain circuit parallelism. In \cite{graham_multi-qubit_2022}, any local \texttt{R}$(\theta,\phi)$ gates with different $\theta$ or $\phi$ are decomposed using \texttt{GR} gates with different parameters. These are therefore forced to be serialized, even if the \texttt{R}$(\theta,\phi)$ gates were originally scheduled in the same moment, resulting in exceedingly high circuit depths and global gate counts. 

We next compare the Transverse decomposition and $\theta$-Opt scheduling combination relative to the Axial and \texttt{stratified\_circuit} combination. We observe up to 4.77x speedup, with a geometric mean of 3.02x speedup. Up to 2.85x of this speedup is from the Transverse decomposition, and up to 1.75x from $\theta$-Opt. These results are shown in Fig.~\ref{fig: results circuit duration speedup}.

\begin{figure}
    \includegraphics[scale=0.365]{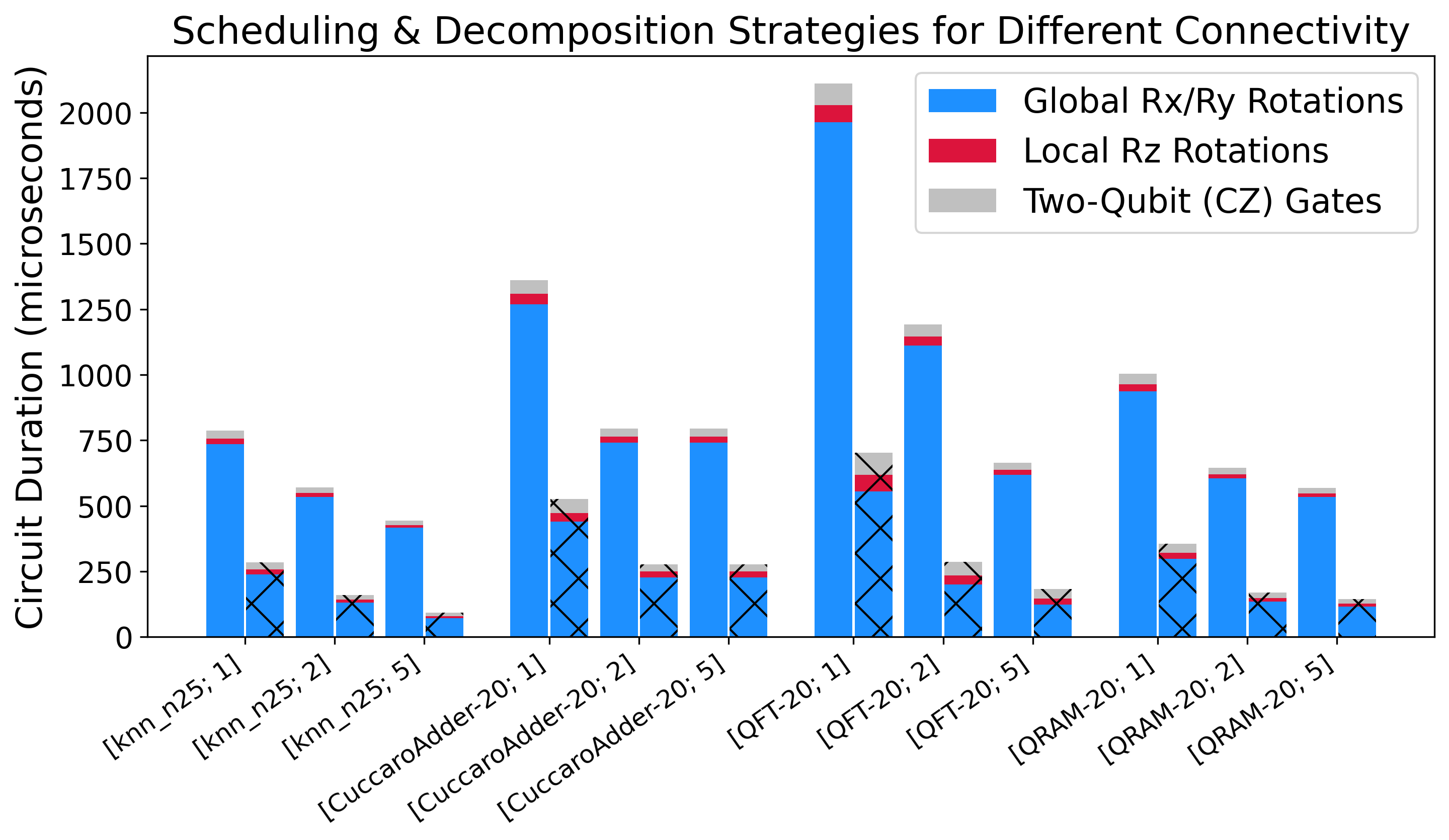}
    \caption{Impact of varying connectivity, as defined by the ratio between blockade radius and atom spacing. Bars are labeled by the benchmark, number of qubits, and connectivity. Unhatched bars show circuit duration for \texttt{stratified\_circuit} + Axial, while hatched bars show $\theta$-Opt + Transverse. Similar patterns are observed with all other benchmark circuits.}
    \label{fig: results connectivity}
\end{figure}

In Fig.~\ref{fig: results connectivity}, circuit duration is broken down by gate type, showing that the majority of time is spent executing global gates. The benefit gained from the Transverse decomposition and $\theta$-Opt scheduling comes from specifically reducing this cost. For some benchmarks\textemdash particularly those that have a large portion of \texttt{U3} gates with $\theta\in\set{\pi/2,\pi}$, e.g., the Cuccaro Adder, QFT Adder, or CNX Log Depth circuits from \cite{qcb}\textemdash the Axial decomposition results in lower \texttt{Rz} rotation angles than the Transverse decomposition. However, because \texttt{Rz} costs make up such a small fraction of the total circuit duration, this has almost negligible impact on the overall speedups. 

For the Transverse decomposition, the circuits with the most speedup are those with \texttt{U3} gates that have, on average, lower $\theta$ values. This allows for lower \texttt{GR} rotation angles when Eq.~\ref{eqn: transverse full moment}-\ref{eqn: transverse kappa}  are applied, as compared to the set $\pi/2$ rotation angle used with the Axial decomposition. For the $\theta$-Opt scheduling, we see the greatest improvement in circuits with the most flexibility to move gates between different moments, based on dependencies. We find this to be correlated to circuit parallelism, which can be quantified as a number from 0 to 1 by the parallelism factor $f_\parallel$ in \cite{tomesh_supermarq_2022}. In particular, circuits with either very high parallelism (e.g., \texttt{dnn\_16} \cite{li_qasmbench_2022}, $f_\parallel=0.914$) or very low parallelism (e.g., \texttt{cat\_state\_n22} \cite{li_qasmbench_2022}, $f_\parallel=0$) benefit the least from $\theta$-Opt. Of the benchmarks we test, the greatest speedup comes from \texttt{knn\_n25} \cite{li_qasmbench_2022}, which has $f_\parallel=0.605$. We note that $\theta$-Opt is guaranteed to never give a worse schedule than the \texttt{stratified\_circuit} baseline.

\subsection{Improvement in Circuit Fidelity}\label{section: results fidelity}

  Our compiler provides up to $10^7$x improvement in circuit fidelity when comparing the Transverse and $\theta$-Opt combination to Axial and \texttt{stratified\_circuit}, as shown in Fig.~\ref{fig: results fidelities}. Relative to the methods in \cite{graham_multi-qubit_2022}, we see even greater fidelity improvement. As explained in Sec.~\ref{section: fidelity calculations}, larger rotation angle leads to lower gate fidelity. Additionally, idle errors depend indirectly on gate angles, since gate duration is proportional to rotation angle. Therefore, as in Sec.~\ref{section: results circuit duration}, the majority of this improvement is due to minimizing \texttt{GR} rotation amount. 

With fidelity comparisons, circuits with the largest gate \textit{counts} see the most improvement; e.g., the CNX In Place and Select Circuits from \cite{qcb} have significantly larger gate counts relative to the other circuits with similar number of qubits. This is because gate errors dominate over idle errors (which are more closely correlated with circuit \textit{depth}) on neutral atoms devices, and the higher the gate count in the circuit, the more those angle-dependent errors will accumulate if rotation amount is not optimized. For circuits with smaller gate counts, this may not result in a huge difference. However, as circuit sizes scale up, our optimizations become increasingly crucial. 

\begin{figure}[t!]
    \centering

    \begin{subfigure}[t]{0.97\columnwidth}
        \centering
        \includegraphics[width=\textwidth]{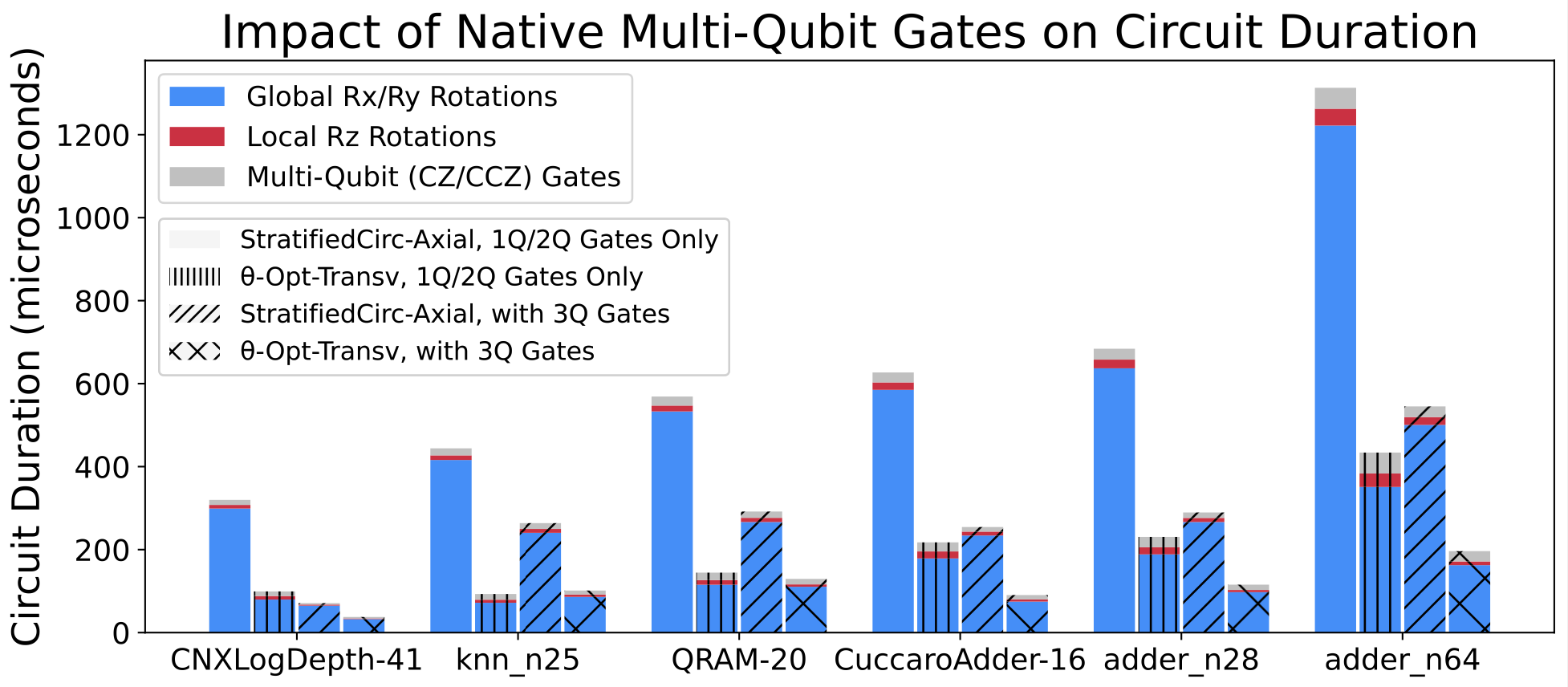}
        \caption{
        }
        \label{fig: results multiqubit gates durations}
    \end{subfigure}\hspace{0.05\textwidth}
        ~ 
    \begin{subfigure}[t]{0.96\columnwidth}
        \centering
        \includegraphics[width=\textwidth]{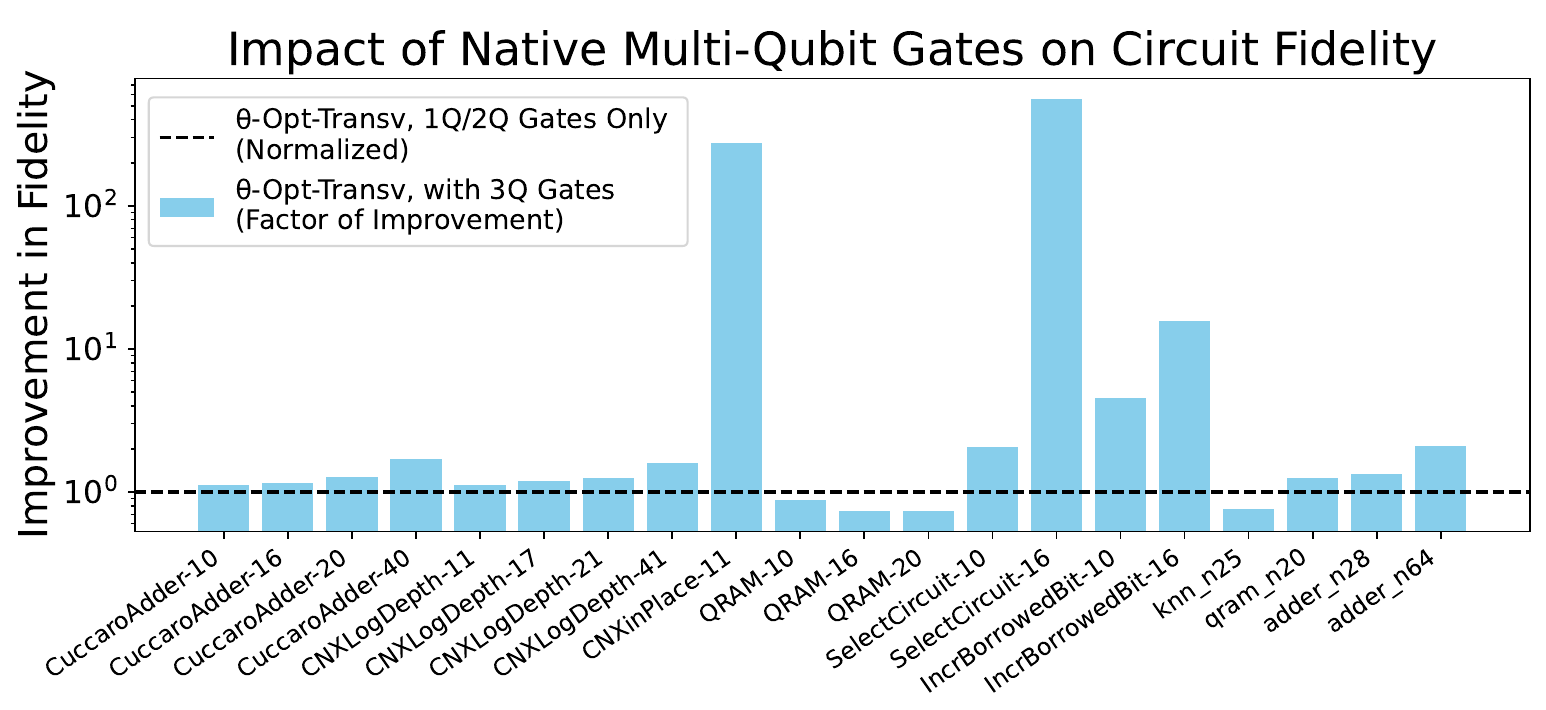}
        \caption{
        }
        \label{fig: results multiqubit gates fidelities}
     \end{subfigure}
    \caption{(a) Circuit duration comparisons when compiled with 3-qubit gates vs. without 3-qubit gates, for both the \texttt{stratified\_circuit} + Axial combination and the $\theta$-Opt + Transverse combination. The benchmarks that we show here are representative of the results we see with all other benchmarks that contain 3-qubit gates in the original circuit. (b) Factor of improvement in fidelity that is gained when native three-qubit gates are incorporated, using $\theta$-Opt + Transverse in both cases. Bars below the black dashed line indicate that the circuit fidelity is better when compiled using only 1-qubit and 2-qubit gates. For these experiments, we use the routing method in \cite{baker_exploiting_2021}, which accounts for multi-qubit gates.}
    \label{fig: results multiqubit gates}
\end{figure}

\subsection{Impact of Varying Connectivity}

One advantage with neutral atoms is their long-range interactions, allowing for higher connectivity. Our goal in this subsection is to understand 1) how much benefit we achieve by increasing connectivity, and 2) if the results described in the previous subsections---i.e., which compiler methods perform the best and by how much---remain the same.  

We define connectivity as the maximum number of atom sites between two qubits that can interact via entangling gates, given by the ratio of blockade radius to atom spacing. Greater connectivity means lower costs due to routing, but it also results in less parallelism, since gates acting on atoms with overlapping blockade radii must execute in serial. Fig.~\ref{fig: results connectivity} demonstrates that the benefit of the former outweighs the extra costs of the latter, and that this benefit plateaus beyond connectivity of 2 to 3. Furthermore, we observe that connectivity does not change the factor of improvement achieved from our optimizations, both with circuit duration and fidelity.

\subsection{Impact of Incorporating Native Multi-Qubit Gates}

Another advantage of neutral atom platforms is the ability to natively implement multi-qubit entangling gates such as \texttt{CCZ}. Though each individual \texttt{CCZ} has longer duration and lower fidelity than any individual single-qubit or two-qubit gate we consider, incorporating multi-qubit gates results in lower circuit depths and gate counts\textemdash often leading to better duration and fidelity for the entire circuit. We explore this tradeoff in Fig.~\ref{fig: results multiqubit gates} for benchmarks with three-qubit gates (e.g.,~\texttt{CSWAP} or Toffoli gates) in the original input circuit, finding that the majority of circuits see a net benefit when \texttt{CCZ} gates are included in the final gate set. The exception is the \texttt{knn\_n25} \cite{li_qasmbench_2022} and \texttt{QRAM} circuits \cite{qcb}, which\textemdash due to a less significant reduction in circuit depth and gate count when \texttt{CCZ} gates are used\textemdash have higher circuit fidelity when the three-qubit gates are decomposed into single-qubit and two-qubit gates. 

\section{Conclusion}

We present the first systems-level work aimed at overcoming the challenges of limited local addressability in many current neutral atom architectures. Our compiler translates input circuits into a native gate set involving global gates, and our decomposition and scheduling passes specifically minimize the global gate rotation amount in the final circuit. These optimizations are crucial in reducing the large gate counts, circuit depths, and error accumulation that would otherwise occur with decompositions involving global gates, thereby significantly improving circuit durations and circuit fidelities.

Though local addressing of all single-qubit gates is possible, and though this may be adopted by more architectures in the future, local addressability comes with greater hardware challenges. Systems-level optimizations, such as those presented here, may reduce the need for higher-complexity hardware solutions. These software optimizations are essential to fully understand the tradeoffs between global vs.~local gates. We note that these tradeoffs may look different in a NISQ vs.~error correction setting. As of now, it is unclear the role that global vs.~local addressability will play within an error correction context, and future work will study this in greater depth.

\section*{ACKNOWLEDGMENTS}
This work is funded in part by EPiQC, an NSF Expedition in Computing, under award CCF-1730449; in part by STAQ under award NSF Phy-1818914/232580; in part by the US Department of Energy Office of Advanced Scientific Computing Research, Accelerated 
Research for Quantum Computing Program; in part by the NSF Quantum Leap Challenge Institute for Hybrid Quantum Architectures and Networks (NSF Award 2016136), in part based upon work supported by the U.S. Department of Energy, Office of Science, National Quantum 
Information Science Research Centers, and in part by the Army Research Office under Grant Number W911NF-23-1-0077; and in part by the National Science Foundation Graduate Research Fellowship under Grant No.~2140001. The views and conclusions contained in this document are those of the authors and should not be interpreted as representing the official policies, either expressed or implied, of the U.S. Government. The U.S. Government is authorized to reproduce and distribute reprints for Government purposes notwithstanding any copyright notation herein.

FTC is the Chief Scientist for Quantum Software at Infleqtion and an advisor to Quantum Circuits, Inc.

\bibliographystyle{IEEEtran}
\bibliography{refs}

\newpage
\appendix

Here, we derive Eq.~\ref{eqn: transverse full moment}-\ref{eqn: transverse kappa} that are used in the Transverse decomposition, discussed in Sec.~\ref{section: transverse decomposition} in the main text.

\subsubsection{Derivation of $\chi$}

Suppressing the qubit index $j$, the key to deriving Eq.~\eqref{eqn: transverse full moment} is finding angles $(\chi,\alpha,\beta)$ for which $A=B$, where
\begin{align}
  A = \texttt{Ry}(\theta),
  &&
  B = \texttt{Rz}(\delta_-)
  \texttt{Rv}\left(\chi, \frac{\theta_\tmax}{2}\right)
  \texttt{Rz}(\delta_+),
  \label{eq:L_and_R}
\end{align}
with $\delta_\pm = -(\alpha\pm\beta)$.
We first seek an angle $\chi$ for which $A$ and $B$ rotate the state $\ket{0}$ to the same latitude.
To this end, we compute
\begin{align}
  p_0
  = \abs{\braket{0|B|0}}^2 = \cos(\chi/2)^2 + \cos(\theta_\tmax/2)^2 \sin(\chi/2)^2.
  \label{eq:prob_0_chi}
\end{align}
Expanding $\cos(x)^2 = 1 - \sin(x)^2$, we can find that
\begin{align}
  \sin(\chi/2)^2 = \frac{1-p_0}{1-\cos(\theta_\tmax/2)^2},
  \label{eq:sin_chi}
\end{align}
and similarly
\begin{align}
  \cos(\chi/2)^2 = \frac{p_0 - \cos(\theta_\tmax/2)^2}{1-\cos(\theta_\tmax/2)^2}.
  \label{eq:cos_chi}
\end{align}
Dividing Eq.~\eqref{eq:sin_chi} by Eq.~\eqref{eq:cos_chi} and substituting $p_0 = \abs{\braket{0|A|0}}^2 = \cos(\theta/2)^2$, we thus find
\begin{align}
  \tan(\chi/2)^2 = \frac{1 - \cos(\theta/2)^2}{\cos(\theta/2)^2 - \cos(\theta_\tmax/2)^2},
  \label{eq:tan_chi}
\end{align}
which is equivalent to Eqs.~\eqref{eqn: transverse chi} and \eqref{eqn: transverse kappa} (tentatively ignoring the sign $\sigma_j=\pm1$).
Note that real-valued angles $\chi$ satisfying Eq.~\eqref{eq:tan_chi} only exist when $\abs{\theta}\le\abs{\theta_\tmax}$.

%%%%%%%%%%%%%%%%%%%%%%%%%%%%%%%%%%%%%%%%%%%%%%%%%%
\subsubsection{Derivation of $\alpha$ and $\beta$}

With foresight, we now move the \texttt{Rz} gates in Eq.~\eqref{eq:L_and_R} to the other side of $A=B$ and define
\begin{align}
  \tilde{A} = \texttt{Rz}(-\delta_-)
  \texttt{Ry}(\theta)
  \texttt{Rz}(-\delta_+),
  &&
  \tilde{B} = \texttt{Rv}\left(\chi, \frac{\theta_\tmax}{2}\right),
\end{align}
so that $\tilde{A} = \tilde{B}$.
We define the inner product $\braket{P,O} = \Tr(P^\dag O)/2$, and denote the single-qubit Pauli operators (with identity) by $I,X,Y,Z$.
We then compute
\begin{align}
  \frac{\braket{Z,\tilde{A}}}{\braket{I,\tilde{A}}}
  = -i \tan\alpha
  = \frac{\braket{Z,\tilde{B}}}{\braket{I,\tilde{B}}}
  = -i \cos(\theta_\tmax/2) \tan(\chi/2),
  \label{eq:tan_alpha}
\end{align}
which implies $\tan\alpha = \cos(\theta_\tmax/2) \tan(\chi/2)$, or equivalently Eq~\eqref{eqn: transverse alpha}.
Looking at $Y$-components of $\tilde{A}$ and $\tilde{B}$,
\begin{align}
  \braket{Y,\tilde{A}}
  = -i \cos\beta \sin(\theta/2)
  = \braket{Y,\tilde{B}}
  = 0,
  \label{eq:cos_beta}
\end{align}
which tells us that $\beta$ must be a half-integer multiple of $\pi$.
Without loss of generality, we enforce $\beta=\pm\pi/2$.
To resolve the sign ambiguity, we examine $\tilde{A}$ and $\tilde{B}$'s $X$-components:
\begin{align}
  \braket{X,\tilde{A}}
  = \sin\beta \sin(\theta/2)
  = \braket{X,\tilde{B}}
  = \sin(\chi/2) \sin(\theta_\tmax/2).
  \label{eq:sin_beta}
\end{align}
Assuming without loss of generality that $\chi>0$ and $\theta,\theta_\tmax\in(-2\pi, 2\pi)$, we conclude Eq.~\eqref{eqn: transverse beta}.
The relations in Eqs.~\eqref{eq:tan_alpha}--\eqref{eq:sin_beta} are invariant under $(\chi,\alpha,\beta)\to(-\chi,-\alpha,-\beta)$, which accounts for the free sign variable $\sigma_j$ in Eqs.~\eqref{eqn: transverse gamma} and \eqref{eqn: transverse chi}.

\end{document}